\documentclass[]{aa}  
\usepackage{graphicx}
\usepackage{txfonts}
\usepackage{amsfonts}
\usepackage{longtable}
\usepackage{pdflscape}
\usepackage{longtable}
\usepackage{afterpage}
\usepackage{changepage}
\usepackage{geometry}
\usepackage{hyperref}
\hypersetup{
    colorlinks=true,
    linkcolor=blue,
    citecolor=blue,
    filecolor=magenta,      
    urlcolor=cyan,
    pdfpagemode=FullScreen,
    }
\usepackage{hhline}
\usepackage{tablefootnote}
\usepackage{natbib}
\bibpunct{(}{)}{;}{a}{}{,}

\usepackage{soul}
\usepackage{color}

\begin{document} 

\title{A new treatment of telluric and stellar features for medium resolution spectroscopy and molecular mapping}
\subtitle{Application to the abundance determination on $\beta$\,Pic\,b}

\author{
F. Kiefer\inst{1} \and     
M. Bonnefoy\inst{2} \and 
B. Charnay\inst{1} \and 
A. Boccaletti\inst{1} \and 
A.-M. Lagrange\inst{1,2} \and
G.~Chauvin\inst{3,2} \and
B.~B\'ezard\inst{1} \and
M.~M\^alin\inst{1} 
}

\institute{
\label{inst:1}LESIA, Observatoire de Paris, Universit\'e PSL, CNRS, Sorbonne Universit\'e, Universit\'e  Paris Cit\'e, 5 place Jules Janssen, 92195 Meudon, France\thanks{Please send any request to flavien.kiefer@obspm.fr} \and
\label{inst:2}Univ. Grenoble Alpes, CNRS, IPAG, F-38000 Grenoble, France \and
\label{inst:3}Laboratoire Lagrange, Université Cote d’Azur, CNRS, Observatoire de la Cote d’Azur, 06304 Nice, France\\
        }

\date{Received -- 05/06/2023 ; accepted -- 11/02/2024}
 
\abstract{Molecular mapping is a supervised method exploiting the spectral diversity of integral field spectrographs to detect and characterize resolved exoplanets blurred into the stellar halo.  We present an evolution of the method to remove the stellar halo and the nuisance of telluric features in the datacubes and access a continuum-subtracted spectra of the planets at R$\sim$4000. We derive planet atmosphere properties from a direct analysis of the planet telluric-corrected absorption spectrum. We applied our methods to the SINFONI observation of the planet $\beta$\,Pictoris\,b. We recover the CO and H$_2$O detections in the atmosphere of $\beta$\,Pic\,b using molecular mapping. We further determine some basic properties of its atmosphere, with $T_\text{eq}$=$1748^{+3}_{-4}$\,K, a sub-solar [Fe/H]=$-0.235^{+0.015}_{-0.013}$\,dex, and a solar C/O=0.551$\pm$0.002 in contrast with values measured for the same exoplanet with other infrared instruments. We confirm a low projected equatorial velocity of 25$^{+5}_{-6}$\,km\,s$^{-1}$. We are also able to measure, for the first time with a medium-resolution spectrograph, the radial velocity of $\beta$\,Pic\,b relative to the central star at MJD=56910.38 with a km/s precision of -11.3$\pm$1.1\,km\,s$^{-1}$, compatible with ephemerides based on the current knowledge of the $\beta$\,Pic system.}


\maketitle

\section{Introduction}

The system of $\beta$\,Pictoris, with its imaged debris disk of dust, evaporating exocomets and two giant planets, is a stunning window on early stages of planetary systems formation and evolution. At the age of $\beta$\,Pic $\sim$23$\pm$3\,Myr~\citep{Mamajek2014}, giant planets have already formed, most of the protoplanetary gas has disappeared from the disk, and Earth-mass planets may be still forming. The discovery of $\beta$\,Pic\,b in direct high-contrast imaging~\citep{Lagrange2009} has been rapidly recognised as a major finding for several reasons. First, it was, until the discoveries of $\beta$\,Pic\,c~\citep{Lagrange2019,Nowak2020b}, and more recently AF Lep b (Mesa et al, 2023), the shortest period imaged exoplanet, allowing thus “fast” orbit characterisation; second, once its mass is known, it can be used to calibrate brightness-mass models and atmosphere models at young ages; third, it is a precious benchmark for detailed atmosphere and physical characterisation thanks to its proximity to Earth and position with respect to the star; and finally, it is an exquisite laboratory to study disk-planet interactions at a post transition disk stage \citep{Lagrange2010,Lagrange2012a}. 

Model and age dependent brightness--mass relationships predict $\beta$\,Pic\,b mass to be within 9-13\,M$_{\text{J}}$~\citep{Bonnefoy2013,Morzinski2015,Chilcote2017}. Its mass is still marginally constrained observationnally, because of significant uncertainties on the amplitude of the radial velocities (RV) variations induced by the planets b \& c. In particular, the available RV data do not cover the whole $\beta$\,Pic\,b period, the extrema of the recently discovered $\beta$\,Pic\,c induced variations are not well constrained with the available data, and the RV variations are strongly dominated by the stellar pulsations (see examples in \citealt{Lagrange2019,Lagrange2020,Vandal2020}). Gaia was used by several authors to further constrain the planet $b$ mass, $<$20\,M$_\text{J}$~\citep{Bonnefoy2013}, 13$\pm$3\,M$_\text{J}$~\citep{Dupuy2019}, 12.7$\pm$2.2\,M$_\text{J}$~\citep{Nowak2020}, 10-11\,M$_\text{J}$~\citep{Lagrange2020}. The most recent determinations combine RV, relative and absolute astrometry, taking into account both planets b \& c. They lead to $9.3_{-2.5}^{+2.6}$\,M$_\text{J}$ using the Hipparcos--GaiaDR2 measurement of astrometric acceleration~\citep{Brandt2021}, and 11.7$^{+2.3}_{-2.1}$\,M$_\text{J}$ using the Hipparcos--GaiaDR3 measurement of astrometric acceleration with the same datasets~\citep{Feng2022}. We note that the astrometric acceleration measurement, also known as proper motion anomaly (see also~\citealt{Kervella2019,Kervella2022}), initially 2.54--$\sigma$ significant using the DR2~\citep{Kervella2019}, became compatible with zero at 0.86--$\sigma$ using the DR3~\citep{Kervella2022}. This explains the difference in the derived mass. From dynamical considerations, the mass of $\beta$\,Pic\,b is thus bounded within 9--15\,M$_\text{J}$.

The study of infrared spectra emitted by transiting and non-transiting hot Jupiter~\citep{Brogi2012,Brogi2013,deKok2013,Birkby2013,Lockwood2014,Brogi2014,Piskorz2016,Piskorz2017,Birkby2017,Guilluy2019,Cont2021,Cont2022a,Yan2022,Cont2022b} and young imaged planets~\citep{Snellen2014,Brogi2018,Hoeijmakers2018,Petit2018,Ruffio2019,Nowak2020,Ruffio2021,Cugno2021,Petrus2021,Patapis2022,Malin2023,Petrus2023,Miles2023,Landman2023} allows us to characterize the atmospheric composition in molecules such as CO, CO$_2$, H$_2$O, NH$_3$ and CH$_4$. The molecular mapping method was first developed for this objective by \citet{Snellen2014}, hereafter S14, for medium or high resolution instruments such as CRIRES (S14,~\citealt{Landman2023}), SINFONI~\citep{Hoeijmakers2018,Cugno2021,Petrus2021}, Keck/OSIRIS~\citep{Petit2018,Ruffio2019,Ruffio2021} or JWST/MRS~\citep{Patapis2022,Malin2023,Miles2023}. This method consists in calculating the cross-correlation function (CCF) of a spectrum emitted from the atmosphere of a planet with a theoretical transmission spectrum, or template, using for instance Exo-REM~\citep{Baudino2015,Charnay2018}. This could reveal the presence of individual molecules. The CCF leads to a similarity score, which if equal to 1 (0) means the spectrum and the template are proportional (totally orthogonal). In general because of noise, systematics, and inaccuracies of models, a CCF never reaches exactly 1. Using the CCF as a template matching score in principle allows us to retrieve simple atmospheric properties such as $T_\text{eff}$, $\log g$ and relative abundances. 

For $\beta$\,Pic\,b, \citet{Hoeijmakers2018} (H18 hereafter) showed using molecular mapping on the cubes collected by the Spectrograph for INtegral Field Observations in the Near Infrared (or SINFONI) that H$_2$O and CO were present in the atmosphere of this young planet, with no evidence of other species. They did a tentative template matching of $T_\text{eff}$ and $\log g$ that led only to large confidence regions of those parameters on their Figure 10. They did not produce any estimation of the planet radial velocity nor rotational broadening $v\sin i$. 

With a spectrum-fitting oriented approach, an emitted spectrum of $\beta$\,Pic\,b was obtained with GRAVITY~\citep{Nowak2020}. Its fit led, in a Bayesian inference framework using Markov chain Monte-Carlo sampling of posteriors,  to an effective temperature of 1740$\pm$10\,K with $\log g$=4.35$\pm$0.09, a super-solar metallicity [M/H]$\sim$0.7$\pm$0.1 and a sub-solar C/O=0.43$\pm$0.04\footnote{The commonly adopted solar abundance values are $\log N_{M,\odot}/N_{H,\odot}$=$\log Z_\odot/X_\odot$=-1.74 and C/O=0.55~\citep{Asplund2009}.}. This was in good agreement with previous estimations of the planet temperature of 1724\,K and a $\log g$=4.2 by~\citet{Chilcote2017} and the combined astrometric+RV planet mass estimation $\sim$12\,M$_\text{J}$~\citep{Snellen2018,Nowak2020,Lagrange2020}. However, the metallicity was significantly different if considering the GRAVITY spectrum only (-0.5\,dex) or combined with the GPI YJH low-resolution spectra  ($>$0.5\,dex). Most recently,~\citet{Landman2023} published the analysis of new $\beta$\,Pic\,b high-resolution spectra taken with the upgraded CRIRES+ instrument that led to similar parameters using atmospheric retrievals, with temperatures slightly higher than in~\citet{Nowak2020}, a sub-solar metallicity (Fe/H$\sim$-0.4) and a sub-solar C/O=0.41$\pm$0.04. Allowed by the high-resolution they obtained a new $v\sin i$ measurement of 19.9$\pm$1.1\,km\,s$^{-1}$.

With an approach similar to~\citep{Nowak2020}, \citet{Petrus2021} used both principal component analysis (PCA) and Halo-subtraction on North-aligned angular differential imaging (nADI) on SINFONI observations of HIP\,65426 to extract the emitted spectrum of the planet $b$ keeping the thermal continuum. They then used Bayesian inference with nested sampling~\citep{Skilling2006} to retrieve the basic parameters of planet HIP\,65426\,b from the spectrum itself, including equilibrium temperature, surface gravity, metallicity ratio [M/H] and C/O. This proved possible to derive a spectrum of a planet observed with SINFONI and that having a spectrum-fitting rather than CCF-optimisation method led to more reliable results. 

Here, we perform a new analysis of the $\beta$\,Pic cubes observed with SINFONI. We improve the reduction of the cubes
as thoroughly explained in Section~\ref{sec:sinfoni}. We then improve the star removal method used by H18 with a different approach, that corrects for residuals from stellar lines. We discuss H18's method, and explain our improvements in the form of a new method called \verb+starem+ in Section~\ref{sec:starem}. Then, in Section~\ref{sec:molmap}, we apply molecular mapping and compare to H18 results. We further extract the spectrum of the planet in Section~\ref{sec:average}. We use a simple grid search as well as a Bayesian framework with an MCMC sampling to fit the observed planet spectrum and measure the parameters of the planet. This is done in Section~\ref{sec:template-match}. We discuss the results in Section~\ref{sec:discussion} and give our conclusions in Section~\ref{sec:conclusion}.

\section{SINFONI data pre-processing}
\label{sec:sinfoni}
\subsection{SINFONI observations}

SINFONI was an infrared instrument, coupling an adaptive optics (AO) module to an integral field spectrograph (IFS) SPIFFI, installed on the Unit Telescope 4 of the Very Large Telescope at Paranal/Chile~\citep{Eisenhauer2003,Bonnet2004}. SINFONI was on-sky from 2004 to 2019. The  observations with the SINFONI IFS were performed with different size of field-of-view (FoV) and spectral resolution ($R$), and then reduced into data cubes, with two spatial  and one spectral dimensions. Here, we focus on observations of the $\beta$\,Pictoris surroundings performed with the 0.8\arcsec$\times$0.8\arcsec FoV subdivided into 64$\times$64 spaxels\footnote{Initially there are 64$\times$32 rectangular spaxels of size 12.5$\times$25 mas$^2$, but are then subdivided into 64$\times$64 square spaxels by splitting one spaxel into two equal flux spaxels.} of size 12.5$\times$12.5-mas$^2$ at $R$=4000 along the $K$-band (2.08-2.45\,$\mu$m). The observations consist in 24 exposures of 60\,sec each, recorded on the 10th September 2014 from 08:19:34 UT to 10:05:20 UT. An offset of 0.9-1.1\arcsec from $\beta$\,Pic and a field rotation of -56$^\circ$ to -19$^\circ$ was applied, reducing the pollution of the stellar halo upon the planet spaxels with the star decentered outside the FoV, focusing the observations on the surroundings of $\beta$\,Pictoris, and enabling the use of angular differential imaging~\citep{Marois2006}. The seeing during the observation varied within 0.8-1.0\arcsec with an airmass varying from 1.35 to 1.14 between the first and the last exposure. The atmospheric conditions were relatively constant during the observations with fluctuations of pressure and temperature $<$1\%. 

\subsection{From SINFONI raw data to registered cubes}
\label{sec:textrix}
We performed the data reduction of the SINFONI sequence of observations following the scheme described in~\citet{Petrus2021} which provides optimally-reduced datacubes for high-contrast science.

The raw data were originally corrected the Toolkit for Exoplanet deTection and chaRacterization with IfS (hereafter \texttt{TExTRIS}; \citealt{Petrus2021,Palma2023,Demars2023}, Bonnefoy et al. in prep.) from the so-called "odd-even" effect affecting randomly some pre-amplification channels on SPIFFI's detector (corresponding roughtly to the location of the 25th slitlet). We then used the ESO data handling pipeline version 3.0.0 to reconstruct data cubes from the bi-dimensional science frames.
\texttt{TExTRIS} also corrected for the improper registration of the slitlet edges on the detector and for the inaccurate wavelength solution found by the pipeline using synthetic spectra of the telluric absorptions.

Finally, we used \texttt{TExTRIS} to perform a proper registration of the star position outside the field of view. H18 fitted a synthetic function to represent the wings of the star's point spread function (PSF). However, such a method is sensitive to the distribution of flux within the FoV. The later can be affected by (i) the complex evolution of the Strehl ratio which evolved with wavelength and along the sequence, cubes with high Strehl ratio showing strong artefacts due to the telescope  spiders while those with low Strehl ratio show a smoother flux distribution, and  (ii) the varying part of the halo contained in the FoV due to the field rotation along the sequence.  \texttt{TExTRIS} uses instead an initial measurement of the star position in data cubes acquired during short exposures taken at the beginning of the sequence and centered on the star. Then it builds a model of the $\beta$ Pic centroid positions, which are located outside the FoV in the 24 exposures of the observation sequence, by computing their theoretical wavelength-dependent evolution due to the evolving refraction, the field rotation and the offsets on sky.

A remaining error due to telescope flexure exists (see the ESO user manual) but appears to be below $\sim$1 pixel in the final registered cubes of $\beta$ Pictoris. This reduction provides us with 24 data cubes and associated measurements of the offsets and rotation angles that will be used in Section~\ref{sec:stack} to de-rotate and stack the cubes aligned on the position of the planet $\beta$\,Pic\,b.


\subsection{Reference stellar spectrum}
\label{sec:starspec}

\begin{figure}
    \centering
    \includegraphics[width=89.3mm]{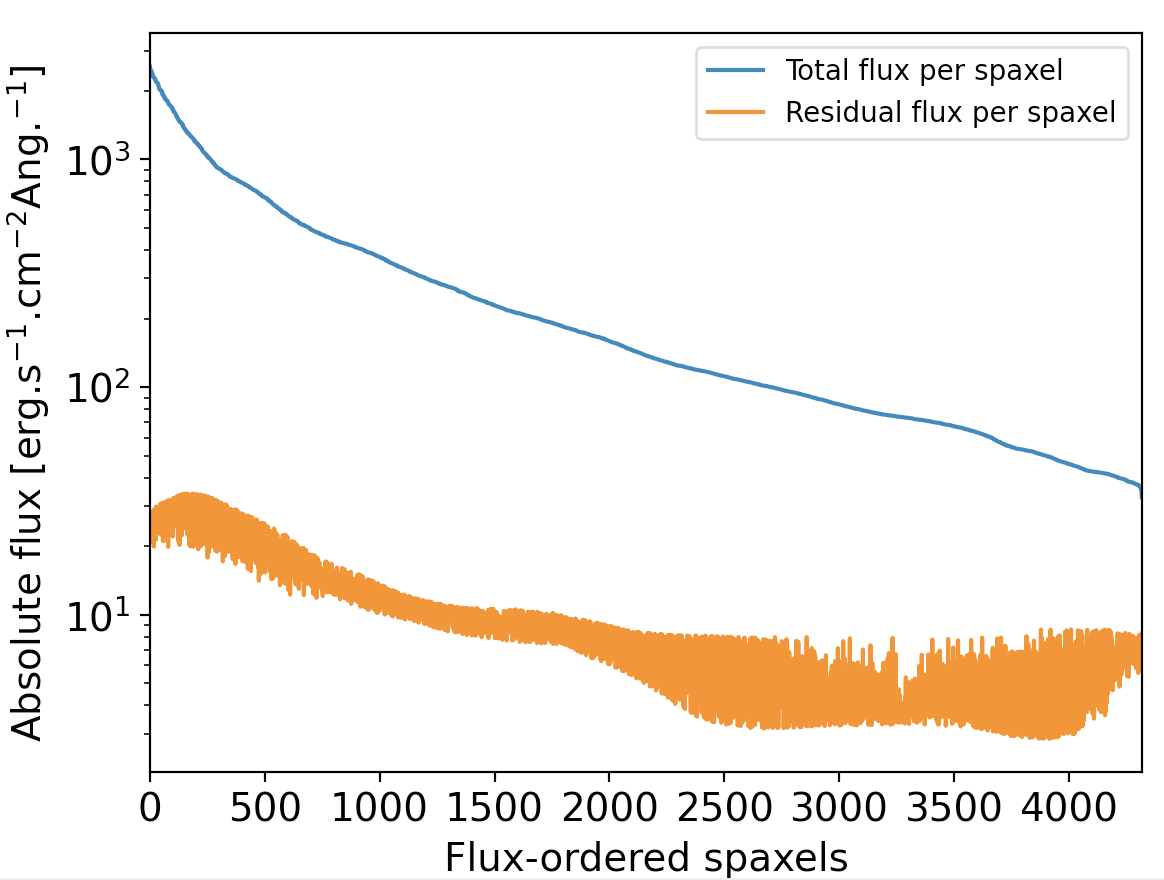}
    \caption{Absolute total (blue) and residual (orange) flux per spaxel. The spaxels are ordered from the brightest to the faintest in absolute total flux per spaxel. The absolute total flux is the raw flux obtained as output of the stack phase in Section~\ref{sec:stack}. The residual flux is the remaining absolute flux once the star spectrum is removed (see Section~\ref{sec:starem} for more details).}
    \label{fig:hist_flux}
\end{figure}

\begin{figure}
    \centering
    \includegraphics[width=89.3mm]{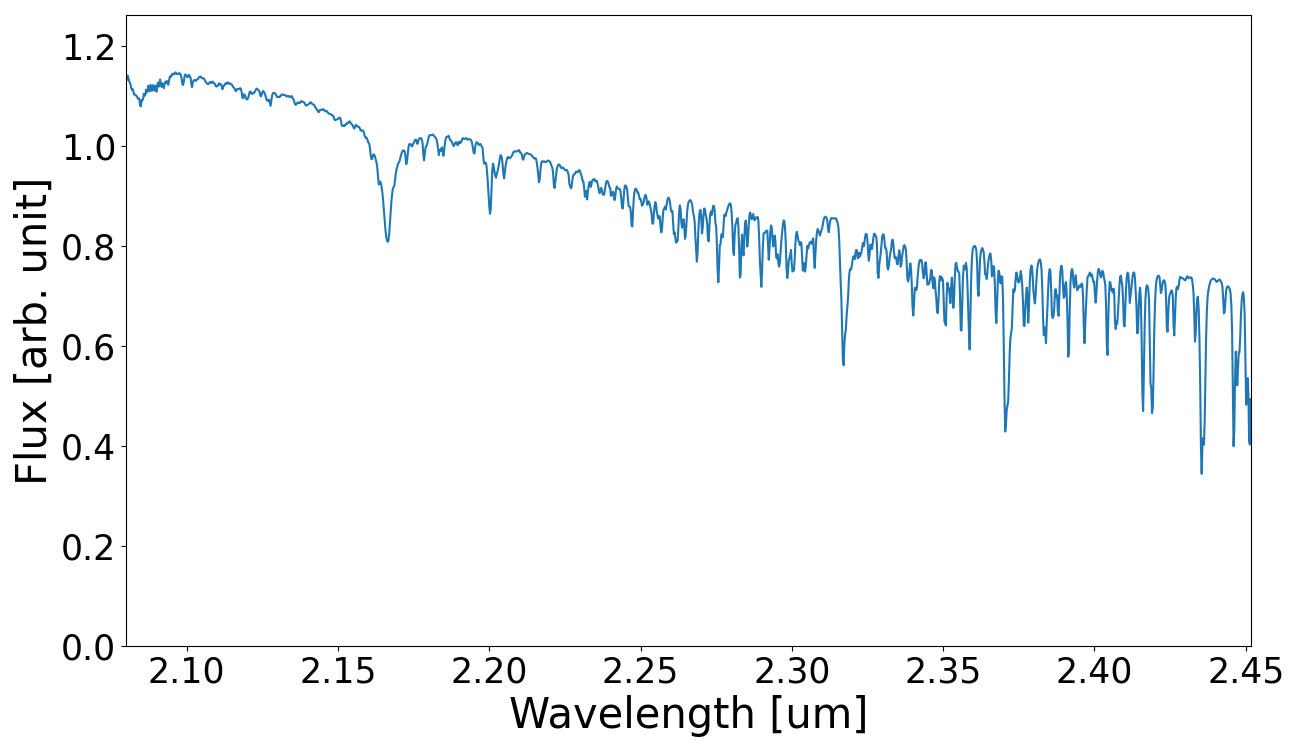}
    \caption{Star spectrum calculated from the brightest spaxels. The flux is normalised to the pseudo-continuum flux at the top of the Br-$\gamma$ line at 2.165\,$\mu$m.}
    \label{fig:star_spectrum}
\end{figure}

As it will be used thoroughly in the rest of the study, we define here the method to derive a reference stellar spectrum, free from photons coming from the planet, in the K-band using a SINFONI data cube. We found best to use several of the brightest spaxels within a cube and combine them to obtain a stellar spectrum, in order to reduce any pollution from the background and the planet. To find those, we measure for all spaxels of the FoV the flux at the continuum-level of the Br-$\gamma$ line at $\sim$2.165\,$\mu$m by fitting the wings of the line by a 2-degree polynomial and retrieving the level of the continuum at 2.165\,$\mu$m. From this flux map, we exclude the 10 brighest spaxels to avoid bad spaxels and calculate an average star spectrum from the next 100 brightest ones. Those are the less affected by the background whose level is on the order of $\sim$20\,erg\,s$^{-1}$\,cm$^{-2}$\,\AA$^{-1}$ while the total flux reaches more than 2000\,erg\,s$^{-1}$\,cm$^{-2}$\,\AA$^{-1}$. Therefore, its contribution is less than 1\% in those spaxels. Fig.~\ref{fig:hist_flux} shows the absolute total and residual flux distribution among spaxels of the stacked cube obtained in Section~\ref{sec:stack} below. 
The resulting reference star spectrum is showed on Fig.~\ref{fig:star_spectrum}. Note that it includes many telluric lines beyond 2.18\,$\mu$m, mainly H$_2$O, CO$_2$ and CH$_4$ lines.

\subsection{Wavelength calibration correction}
\label{sec:calib}
The presence of telluric lines is a nuisance to the analysis of stellar and exoplanet spectra. Nonetheless, they can also be used to adjust the wavelength calibration in the SINFONI cubes. Tellurics can be fitted directly in each of the 24 cubes to the normalised 
star spectrum. Since we are most interested in the planet, the star spectrum is here obtained 
 from the spaxels located at the position of the planet. The planet position in the derotated cube is determined in Section~\ref{sec:CCF} and its PSF of 4-spaxels FWHM in Section~\ref{sec:starem}. We calculated for each cube the mean stellar spectrum on a circular area of 6--spaxels radius around the planet location. 
 
We used the ESO code \verb+molecfit+~\citep{Smette2015,Kausch2015} \verb+v3.13.6+ that implements LBLRTM to perform the telluric line fit in this spectrum. Typical site parameters during the observations are taken as inputs, such as MJD, Paranal altitude and coordinates, humidity ($\sim$4\%), ambiant pressure ($\sim$740\,hPa), ambiant temperature ($\sim$12$^\circ$C), mirror temperature ($\sim$10.9$^\circ$C) and airmass ($\sec z$$\sim$1.1--1.4). \verb+molecfit+ fits the atmospheric parameters -- such as e.g. water column abundance, pressure, temperature etc. -- as well as a continuum, a Chebychev polynomial wavelength solution, and a line spread function to the observed telluric lines in the observed spectrum. The error bars are fixed to the square root of the flux divided by the normalisation. The observed reduced $\chi^2$ are consistent within 1.3--1.4 all through the cubes time series. We found that the molecules H$_2$O, CO$_2$, CH$_4$ and N$_2$O dominate the model tellurics spectrum in the K-band, whilst CO, NH$_3$, O$_2$ and O$_3$ are always either weak and undetermined or fitted to negligible relative abundance values $<10^{-4}$. To reduce the computation effort, we thus only fit for H$_2$O, CO$_2$, CH$_4$ and N$_2$O column densities. 
A polynomial of degree 6 for the fit of the continuum and of degree 1 for the fit of the wavelength solution was adopted. We fixed the LSF to a Gaussian function allowing its width to vary. The LSF is moreover convolved in \verb+molecfit+ by a 1-pixel (0.00025\,$\mu$m) box to mimic the effect of the slit smearing. 

\begin{figure}
    \centering
\includegraphics[width=89.3mm,clip=true,trim=20pt 50pt 70pt 60pt]{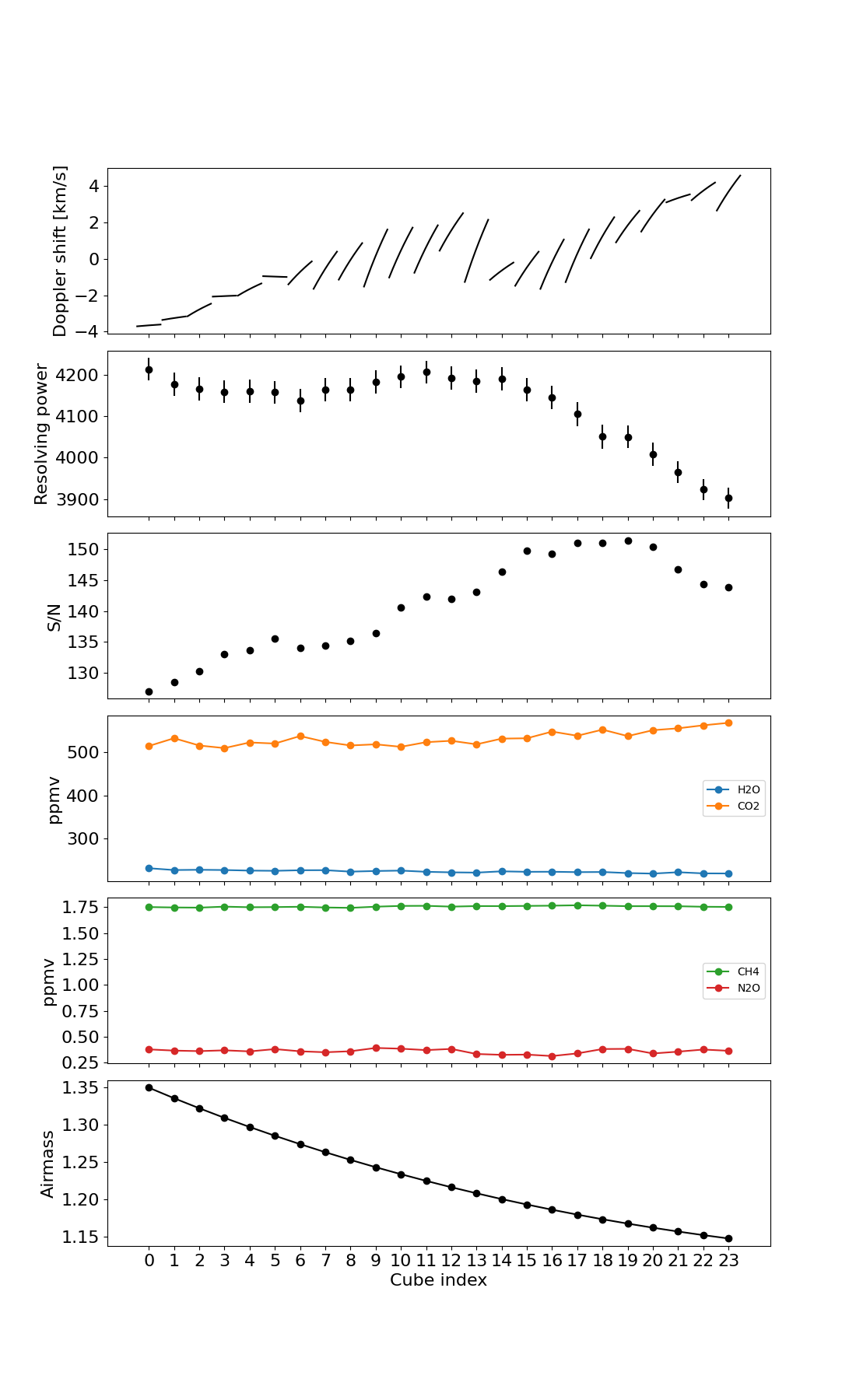}
    \caption{Summary of \texttt{molecfit} results. First panel: wavelength solutions showing the Doppler shift against the wavelength through the K-band with respect to cubes series number. Second panel: the measured resolving power from FWHM of the fitted LSF. Third and fourth panels: fitted ppmv abundances of H$_2$O, CO$_2$, CH$_4$ and N$_2$O.}
    \label{fig:wavelength_sol}
\end{figure}

\begin{figure}
    \centering
    \includegraphics[width=89.3mm]{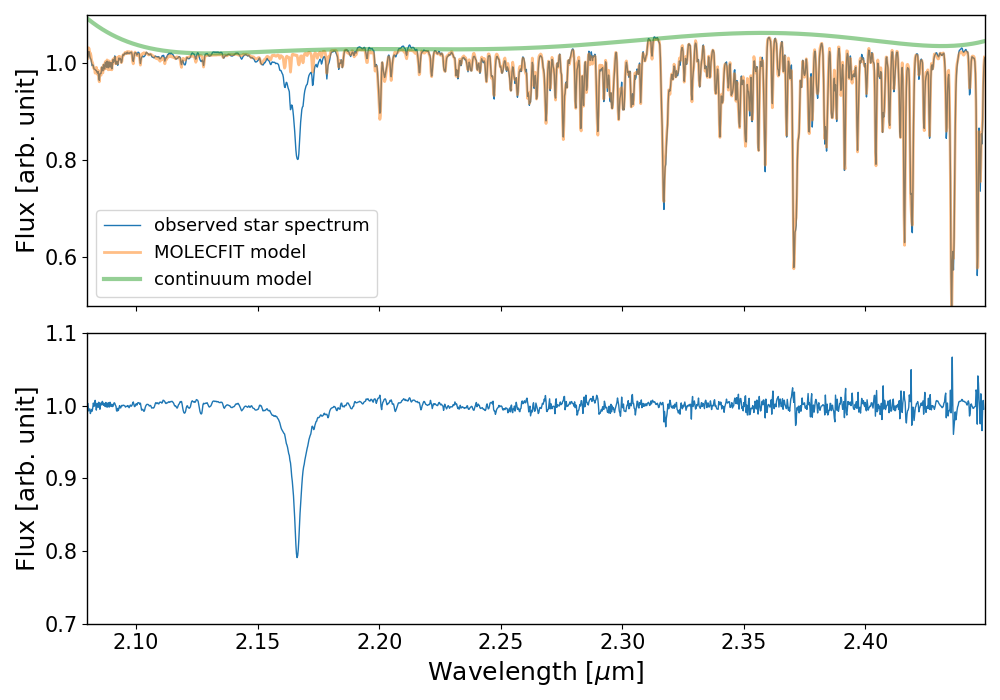}
    \caption{A stellar spectrum observed with SINFONI (blue) compared to the model Earth telluric spectrum calculated with \texttt{molecfit} (orange) and the continuum model (green). Upper panel: the two spectra directly compared. Lower panel: the stellar spectrum then divided by the telluric spectrum model.}
    \label{fig:tellurics}
\end{figure}

Fig.~\ref{fig:wavelength_sol} shows a summary of the \texttt{molecfit} solutions along the cubes. Fig.~\ref{fig:tellurics} shows the $\beta$ Pic stellar spectrum compared with the \texttt{molecfit} resulting model.
There is a residual time-dependent shift of the wavelength solution even after the absolute calibration performed by \texttt{TexTRIS} of about 8 km\,s$^{-1}$ from first (cube 0) to last (cube 23) exposure. This agrees with the magnitude of the error on the calibration found in~\citet{Petrus2021}. Such shift could be due to an effect of flexure of the instrument. We corrected the wavelength solution in all cubes according to this analysis. The figure also shows that the resolving power is varying through the observation series with an LSF FWHM of $\sim$2.16-2.32\,pixels at 2.27\,$\mu$m. This variation is due to the wandering of the planet image on the detector. This leads to an average effective resolving power in our SINFONI K-band spectra of $R$=4120$\pm$90.

The atmospheric molecular column density of H$_2$O is relatively constant along the observation sequence. We note that the CO$_2$ abundance level of $\sim$500\,ppmv retrieved is $\sim$1.4 times higher than the reference value ($\sim$370\,ppmv) in the model Earth atmosphere, while N$_2$O abundance $\sim$0.30\,ppmv is about 1.2 its reference level. At the same date and using the whole K-band spectrum (including deep CO$_2$ and N$_2$O lines below 2.07\,$\mu$m) of another reference star, the CO$_2$ and N$_2$O abundances reach respectively $\sim$390\,ppmv and 0.23\,ppmv (Smette, private comm.). Relying on weak lines only, our determination of the CO$_2$ and N$_2$O abundance levels might not be well determined. The absolute values of the species abundances presented here should thus be considered as indicative only. 

Applying the same procedure on the star spectrum taken from other locations in the image led to similar molecular abundances. However, it revealed a strongly scattered resolving power within 3500--4900, though the average value of $R$ agreed on an effective resolving power of $\sim$4000. This is much smaller than the theoretical $R$=5950, expected for SINFONI in this setup. This might be explained by the degradation due to pixelation when creating the cubes, since the LSF sampling is sub-Nyquist, with a spectral broadening of about 1.5 pixels as noted in the latest SINFONI's documentation\footnote{\texttt{https://www.eso.org/sci/facilities/paranal/\\decommissioned/sinfoni/doc/VLT-MAN-ESO-14700-3517\\-P103.1.pdf}}.

A thorough investigation on how to improve the  effective spectral resolution in the reduced data of instruments such as SINFONI is not in the scope of the present study. Nevertheless, in regards of such objective, two leads might be worth mentioning, that are, either i) achieving a finer reconstruction of cubes from the 2D images of the slitlets and the arc lamp calibration, or more simply, but time-consuming, ii) using dithering during the observations, in order to better sample the LSF, at the cost of at least doubling the exposure time. These leads will be explored in further studies.

\subsection{Science cubes stacking}
\label{sec:stack}
We align the 24 science cubes taken on the 17th of November upon the star centroid. Then, the cubes are de-rotated in such a way that the planet halo is brought at the same ($\alpha^*$,$\delta$)-coordinates in every cube and at every wavelength. We use the values obtained with \texttt{TExTRIS} as explained in Section~\ref{sec:textrix}. This includes a 2D-linear interpolation using the \verb+interpolate.griddata+ routine from \verb+scipy+ in order for all the shifted--rotated cubes to share a common ($\alpha^*$,$\delta$)-grid. Then the 24 cubes are stacked together using a simple average. No clipping of flux is applied during this process in order to maximise the signal-to-noise ratio. This gives us the data cube that we use in the rest of the analysis; we name it the ``master cube'' and it is noted $Q_\text{obs}$.

\section{Star spectrum removal with STAREM}
\subsection{A summary of H18 method}

\begin{figure*}
    \centering
    \includegraphics[width=188.6mm]{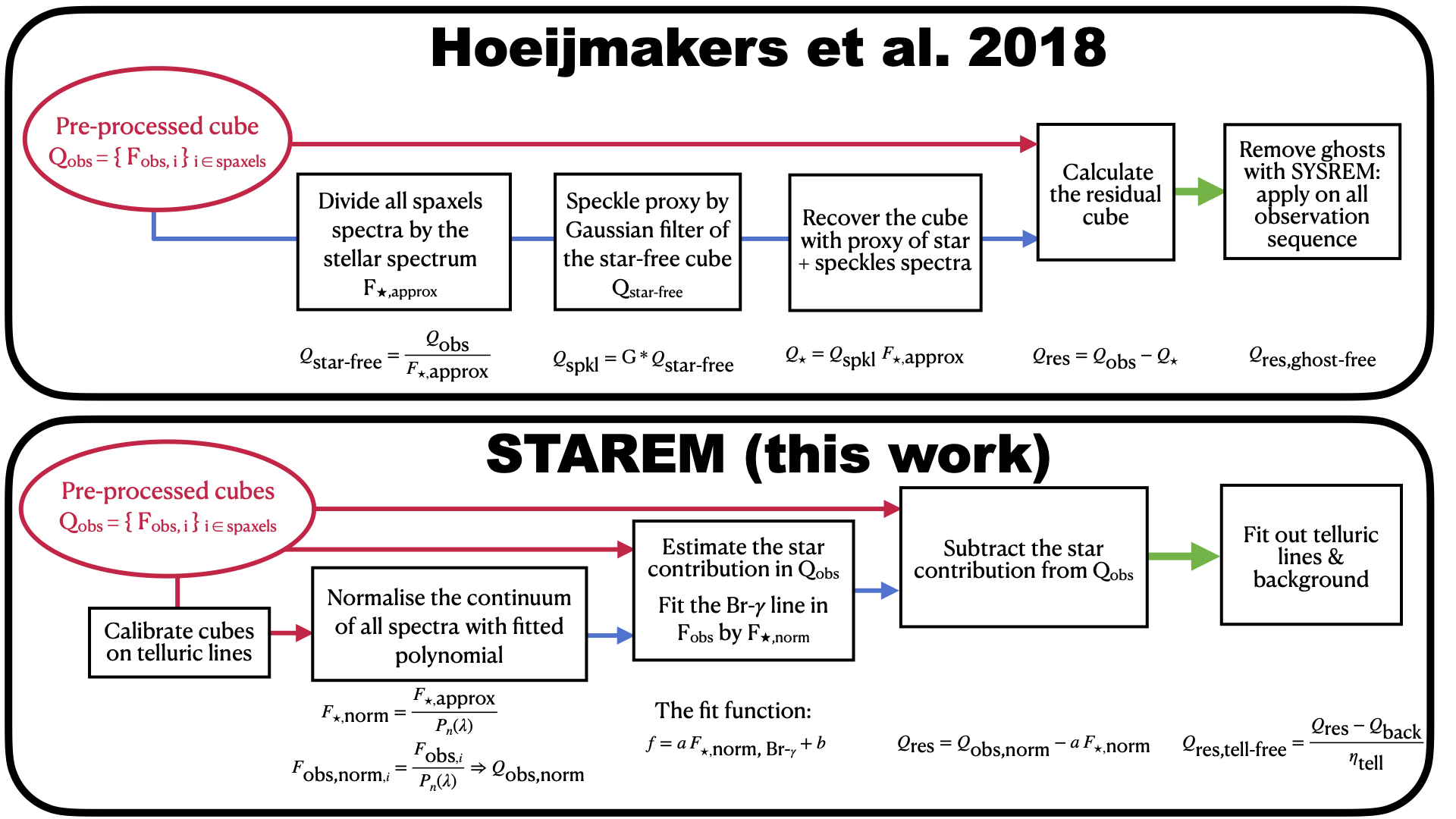}
    \caption{Steps of star spectrum removal methods compared. $Q_\text{obs}$ stands for the stacked cube observed, while $\{Q^{(i)}_\text{obs}\}$ refer to the full series of cubes collected during the night.}
    \label{fig:sketch}
\end{figure*}

The H18 method used to remove the stellar halo proved to work well for performing molecular mapping of exoplanets. However, it does not allow the extraction of a pure planet atmosphere transmission spectrum where systematic deviations remain. We will summarize the H18 method here to show where is the identified issue. A sketch of the different steps of this method is shown in Fig.~\ref{fig:sketch}.

The components of the flux at the planet spaxels $i_p$ are as follows:

\begin{equation} \label{eq:first}
    F_{i_p}(\lambda) = F_{i_p,\star}(\lambda) +  F_{i_p,p}(\lambda) + F_{i_p,B}(\lambda)
\end{equation}

Among which the star flux $F_{i_p,\star}$, the planet flux $F_{i_p,p}$ and a possible background component $F_{i_p,B}$ that we will put aside for the simplicity of the demonstration -- it can be added everywhere by duplicating the planet components and changing 'p' by 'B' in the index -- and will be discussed in Section~\ref{sec:back}. We will also drop the '$i_p$' index in the following to alleviate the equations. 

The planet and star spectra are each composed of a continuum, hereafter noted $C$, multiplied by a 'flat' transmission spectrum whose continuum is normalised to 1 everywhere, hereafter noted $\eta$. Both the astrophysical source and telluric  lines contribute to this transmission spectrum. It can also be expressed as $\eta$=$1-A$, with $A$ a positive comb of spectral lines with the continuum equal to zero everywhere. We point out that $A_{\star}$ and $A_{p}$ for respectively the star and the planet, are supposed to be spaxels-independent since intrinsic to the respective sources. Continua $C_p$ and $C_\star$ are on the other hand spaxel-dependent, because of the point spread function and the wavelength-dependent speckles. 

In order to remove the star and single-out the planet, H18 subtract from each spaxel $i$ of the cube $Q_\text{obs}$ an approximated star spectrum $F_{i,\star,\text{approx}}$ that accounts for wavelength-dependent spaxel-to-spaxel variations. To do so, they divide each spaxel spectrum by the star spectrum determined, as explained in Section~\ref{sec:starspec}, by averaging some of the brightest spaxels of the master cube. The 'star-free' cube $Q_\text{star-free}$ thus obtained show low-frequency wavelength-dependent variations that differ in the spectra from one spaxel to the other. They are due to speckle patterns over the detector changing with wavelength. Those speckle patterns are modeled by applying a Gaussian filter $G$ on this star-free spectrum, $G * Q_\text{star-free}$. It results in a speckle-proxy cube noted $Q_\text{spkl}$ in Fig.~\ref{fig:sketch}. By multiplying those modelled variations by the star spectrum, they finally obtain the star cube $Q_\star$ with at each spaxel $i$ a star spectrum $F_{i,\star,\text{approx}}$  that accounts for wavelength-dependent spaxel-to-spaxel variations.

At any spaxel other than the planet spaxels, subtracting $F_{i,\star,\text{approx}}$ removes the contribution of the star spectrum. However, at the planet spaxel, this is not true. Indeed, the approximated star spectrum contains contributions from both the star and the planet continuum:

\begin{align}
    F_{\star,\text{approx}} =&  \left(C_{p}(\lambda) +  C_{\star}(\lambda)\right) \eta_\star(\lambda)
\end{align} 

By slightly overestimating the star contribution, the subtraction of $F_{\star,\text{approx}}$ rather leads to

\begin{align}
    \Delta F(\lambda) =&  C_{p}(\lambda) \left(A_{\star}(\lambda)-A_{p}(\lambda)\right)
\end{align}

\begin{figure}
    \centering
    \includegraphics[width=89.3mm]{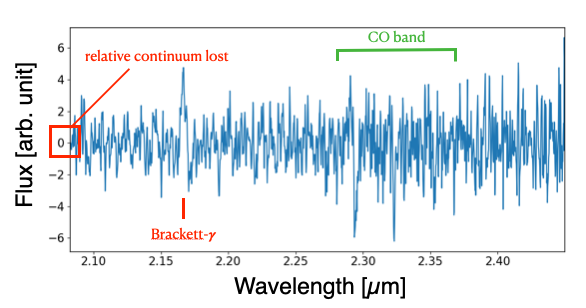}
    \caption{The raw $\beta$\,Pic\,b spectrum extracted from the planet spaxels after using H18 method of star light removal. SYSREM was not used here. We highlight in red the main artifact due to excess star spectrum lines removal at the Brackett-$\gamma$ wavelength. Also should be noticed that the continuum being subtracted, the reference level to which compare the strength of planet molecular lines (CO explicited in green) is missing.}
    \label{fig:hoeijmakers_spec}
\end{figure}

A supplementary stellar contribution -- including tellurics -- to the residual absorption spectrum remains as emission lines with amplitudes comparable to the planet absorption lines. This is shown in Fig.~\ref{fig:hoeijmakers_spec}. Note that the persistence of polluting CH$_4$ lines led H18 to apply several runs of SYSREM~\citep{Tamuz2005} in order to remove them. But this operation does not fix the above issue and stellar lines still remain in the spectrum at the planet spaxels.

The presence of the star spectrum with a non-negligible amplitude is problematic. Moreover, the continuum being subtracted, there are no more reference level to which comparing the strength of molecular lines in the resulting spectrum. As long as the CCF of the star spectrum and the spectra of any of the species found in exoplanet atmosphere is close to zero, this works well for molecular mapping. This is the case here with a 8000\,K star and a $<$2000\,K planet. Still, $\Delta F$ is not strictly-speaking a pure planet atmosphere spectrum. This issue might also explain the detection of an  Br-$\gamma$ emission line in the PDS-70\,b spectrum derived with the same H18 method in~\citet{Cugno2021}: we now suspect it is an artifact from the stellar spectrum removal. 

\subsection{STAREM: a new STAr spectrum REMoval method}
\label{sec:starem}

We propose a different method to subtract the  star spectrum, that we named \texttt{STAREM}, that rather makes use of the normalised transmission spectrum $\eta$=$F/C$. In this spectrum, we are going to estimate the contribution of the star in any spaxel spectrum $F_i$ by fitting stellar absorption lines, and then subtract it from $F_i$. We show this can lead to a well defined flattened transmission spectrum of the planet atmosphere, from which the star spectrum is fully removed, and in which the line strength is preserved. 

First of all, we normalise all spectra of the observed cube, as well as the star spectrum, by fitting a 6th degree polynomial to their continuum, leading to a normalised star spectrum $F_{\star,\text{norm}}$ and a normalised cube $Q_{\text{obs,norm}}$. 
Recalling equation~(\ref{eq:first}), the normalised transmission spectrum components of $F=C\,\eta$ at one of the planet spaxels in $Q_\text{obs}$ are

\begin{equation}
    \eta(\lambda) = \frac{C_{\star}(\lambda) \,\eta_{\star}(\lambda) +  C_{p}(\lambda)\,\eta_{p}(\lambda)}{C(\lambda)}
\end{equation}

We recall that the star and planet pseudo-continua are fixed by the star and planet intrinsic pseudo-continua multiplied by the Strehl ratio and PSF (including speckles) damping of the flux. We introduce the star and planet contribution levels $K_{\star}$ and $K_{p}$ as

\begin{align}
K_{\star}(\lambda)&= 1-K_{p}(\lambda) \label{eq:K}\\
K_{p}(\lambda)&=\frac{C_{p}(\lambda)}{C(\lambda)} \nonumber 
\end{align}.

The transmission spectrum $\eta(\lambda)$ can be expressed more compactly as

\begin{align}
    \eta(\lambda) & = K_{\star}(\lambda) \,\eta_{\star}(\lambda) +  K_{p}(\lambda)\,\eta_{p}(\lambda).
\end{align}

\begin{figure}
    \centering
    \includegraphics[width=89mm,clip=true,trim=15mm 0mm 5mm 15mm]{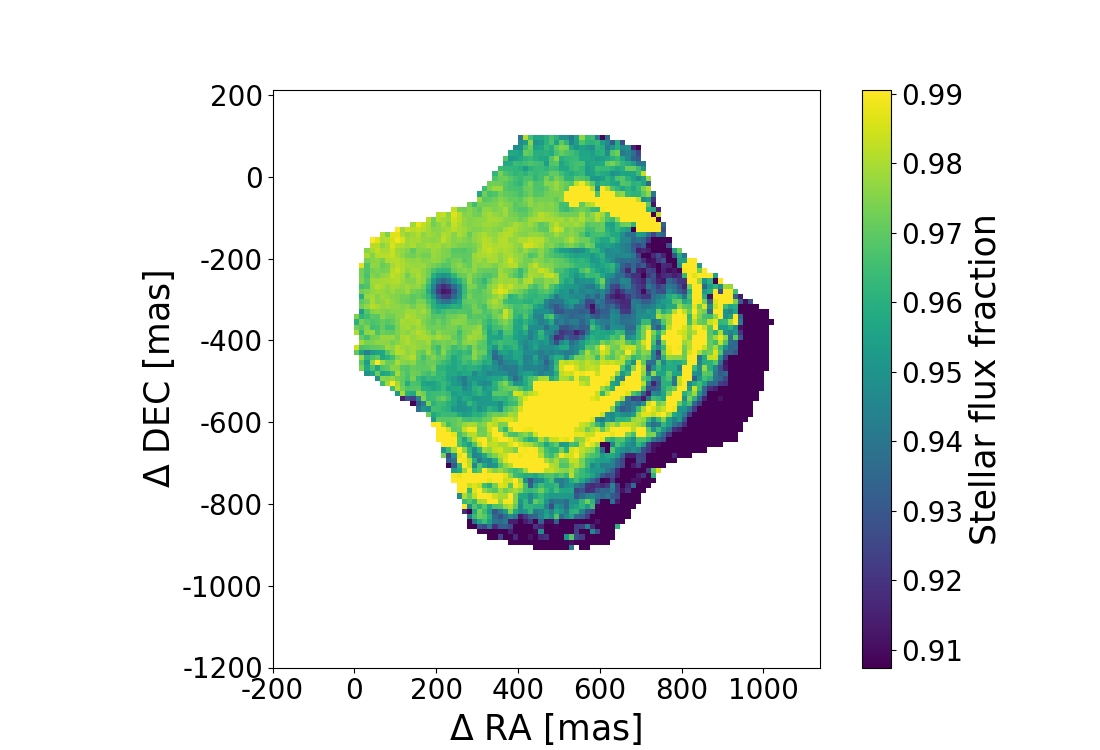}
    \includegraphics[width=89mm,clip=true,trim=15mm 0mm 5mm 15mm]{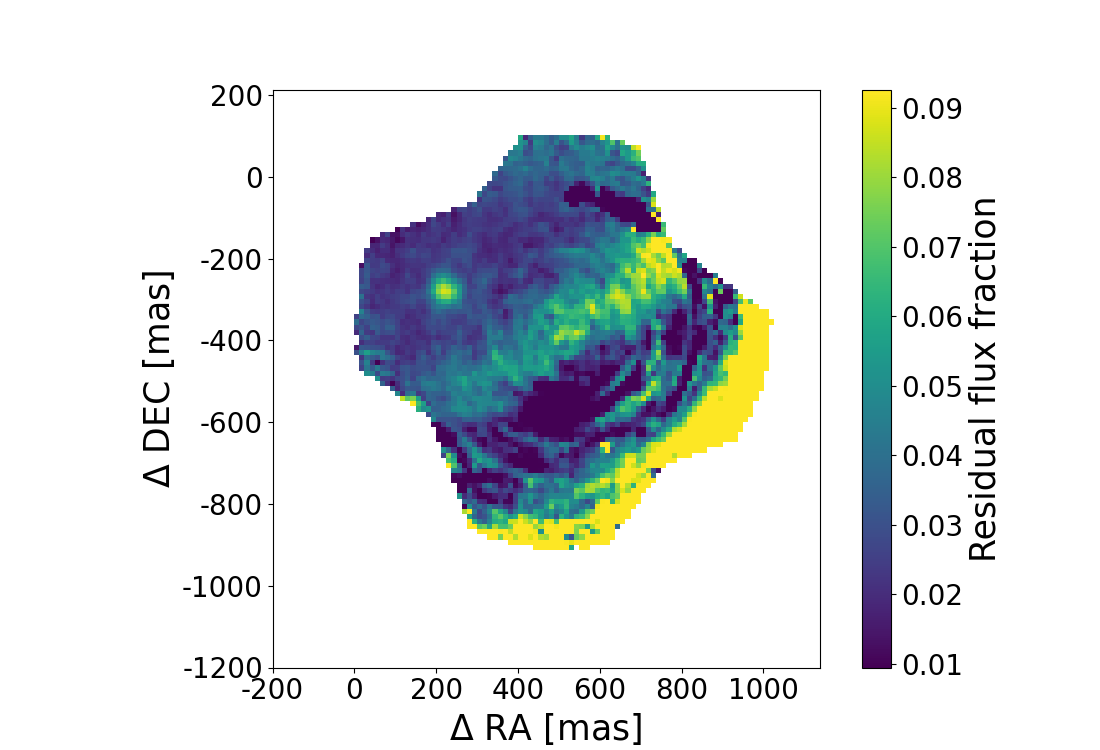}
    \caption{Star flux fraction $K_{i_p,\star}$ at $\lambda$=$2.165\,\mu\text{m}$ (top panel) and the corresponding planet flux fraction 1-$K_{i_p,\star}$ at the same wavelength (bottom panel). It can be seen that some background contributions participate to the residuals.}
    \label{fig:fov_star_contrib}
\end{figure}

Wishing to subtract the star contribution $K_{\star}\,\eta_\star$ from this spectrum, we thus need to estimate $K_{\star}(\lambda)$. This can be achieved by comparing the amplitude of the stellar lines to those of a reference stellar spectrum without contributions from the planet, a ratio of 1 implying a pure stellar spectrum, that is $K_{\star}$=1. Ideally, with several stellar lines present all through the observed band, the best approach would be to use all the lines so as to obtain a more reliable wavelength-dependent approximation of the $K_{\star}(\lambda)$ function. 
In the specific case of the K-band spectrum of $\beta$\,Pic, since the Brackett-$\gamma$ line at 2.165\,$\mu$m is the only strong feature in this spectral band, we are only able to derive an approximated constant star contribution  $\sim$\,\!$K_{\star}(\lambda_0$=$2.165\,\mu\text{m})$, or $K_{\star,2.165}$ hereafter. The derived values of $K_{\star,2.165}$ and its pendent residual contribution, 1-$K_{\star}$, through the SINFONI field-of-view around $\beta$\,Pic\,b are shown in Figure~\ref{fig:fov_star_contrib}.

Removing an approximated contribution $K_{\star,2.165} \,\eta_{\star}$ at each spaxel of $Q_\text{obs,norm}$ leads to a residual cube $Q_\text{res}$. In this cube, the residual spectrum  obtained on a planet spaxel is

\begin{align}
    \eta_\text{res}(\lambda) = & \left(K_{\star}(\lambda) - K_{\star,2.165}\right) \, \eta_{\star}(\lambda) 
    + K_{p}(\lambda)\,\eta_{p}(\lambda)
\end{align}

Using eq.~\ref{eq:K} this could equivalently be written as:
\begin{align}\label{eq:plan_spax}
    \eta_\text{res}(\lambda) = & \left(K_{p,2.165}-K_{p}(\lambda)\right) \, \eta_{\star}(\lambda) 
    + K_{p}(\lambda)\,\eta_{p}(\lambda)
\end{align}

This is a flat spectrum whose baseline level is $K_{p,2.165}$=1-$K_{\star,2.165}$. At the planet spaxels, it should contain the normalised spectrum of the planet with an amplitude corresponding to the relative flux of the planet compared to the star at 2.165\,$\mu$m. 

\begin{figure}
    \centering
    \includegraphics[width=89.3mm]{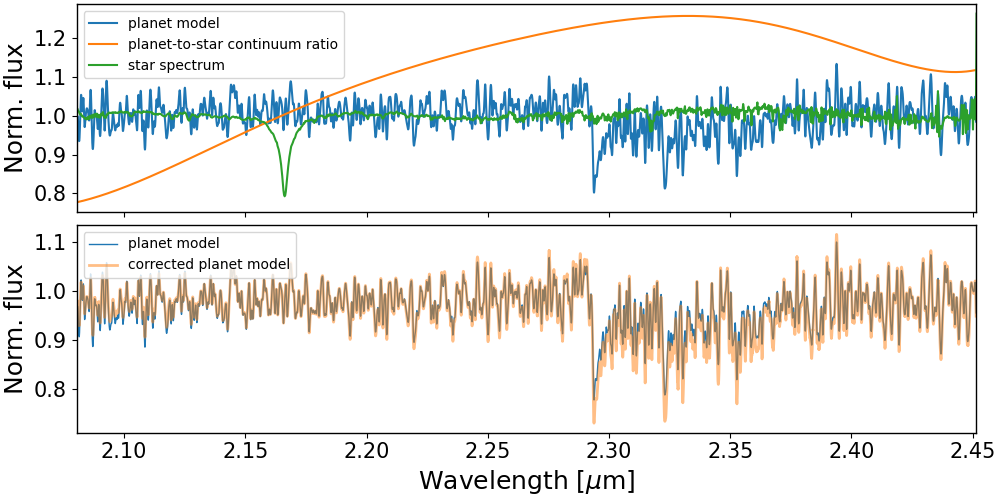}
    \caption{Top panel: the warp function $\delta K (\lambda)$ (orange) normalised to 1 at the Brackett-$\gamma$ wavelength, is compared to a normalised Exo-REM model with $T$=1750K, $\log g$=4.0, $[$Fe/H$]$=0.0 and C/O=0.55 (blue) and the normalised stellar spectrum $\eta_{i_p,\star}$ (green) of $\beta$\,Pic on the K-band. Residual telluric absorptions can be seen in the stellar spectrum. Bottom panel: the naked Exo-REM model (blue) compared to the warped model (orange).}
    \label{fig:ratio_continuum}
\end{figure}

Because of the normalisation of the planet spectra by the continuum of the star, this is however only an approximation. Indeed, away from Brackett-$\gamma$ -- more generally from any fitted stellar line -- the line amplitudes are impacted by a residual star spectrum component with an amplitude $\delta K(\lambda)$=$K_{p,2.165}$-$K_{p}(\lambda)$. This is a footprint of the planet and the star continua within the flat spectrum. Unfortunately, we cannot assume that $\delta K\eta_\star$$\approx $$0$, because it generates a warp of the planet spectrum lines deviating from 1 by a few tens of percent, as shown in Figure~\ref{fig:ratio_continuum}. This will have to be taken into account later on when analysing the spectrum extracted from the planet spaxels.

\subsection{Tellurics and background removal}
\label{sec:back}
In Section~\ref{sec:calib}, tellurics were fitted directly in each of the 24 cubes to the normalised reference star spectrum obtained from the brightest spaxels. They were only used for purpose of calibration. We did not wish to remove tellurics at this step because removal might introduce residuals with amplitudes on the order of magnitude, or worse larger, than the planet features.

Having in hand the residual cube $Q_\text{res}$ that features mostly the planet and background, we now wish to remove the many tellurics lines that are still present in the observed spectrum in the K-band. Most straightfully we could fit a model telluric spectrum to any spaxel spectrum in the cube using \verb+molecfit+. This is however quite time consuming to run \verb+molecfit+ with the cube featuring about 10,000 spaxels, and moreover the S/N of the residual spectra is poor. Instead, we rather use \verb+molecfit+ on the star spectrum derived from the stacked cube at the position of the planet. This allows us to obtain a reliable telluric model $\eta_\text{model}$, to correct again the wavelength solution, using the same settings as in Section~\ref{sec:calib}. Doing so, we determined the effective spectral resolution at the position of the planet in the stacked cube to be 4020$\pm$30. 

\begin{figure}
    \centering
    \includegraphics[width=89.3mm]{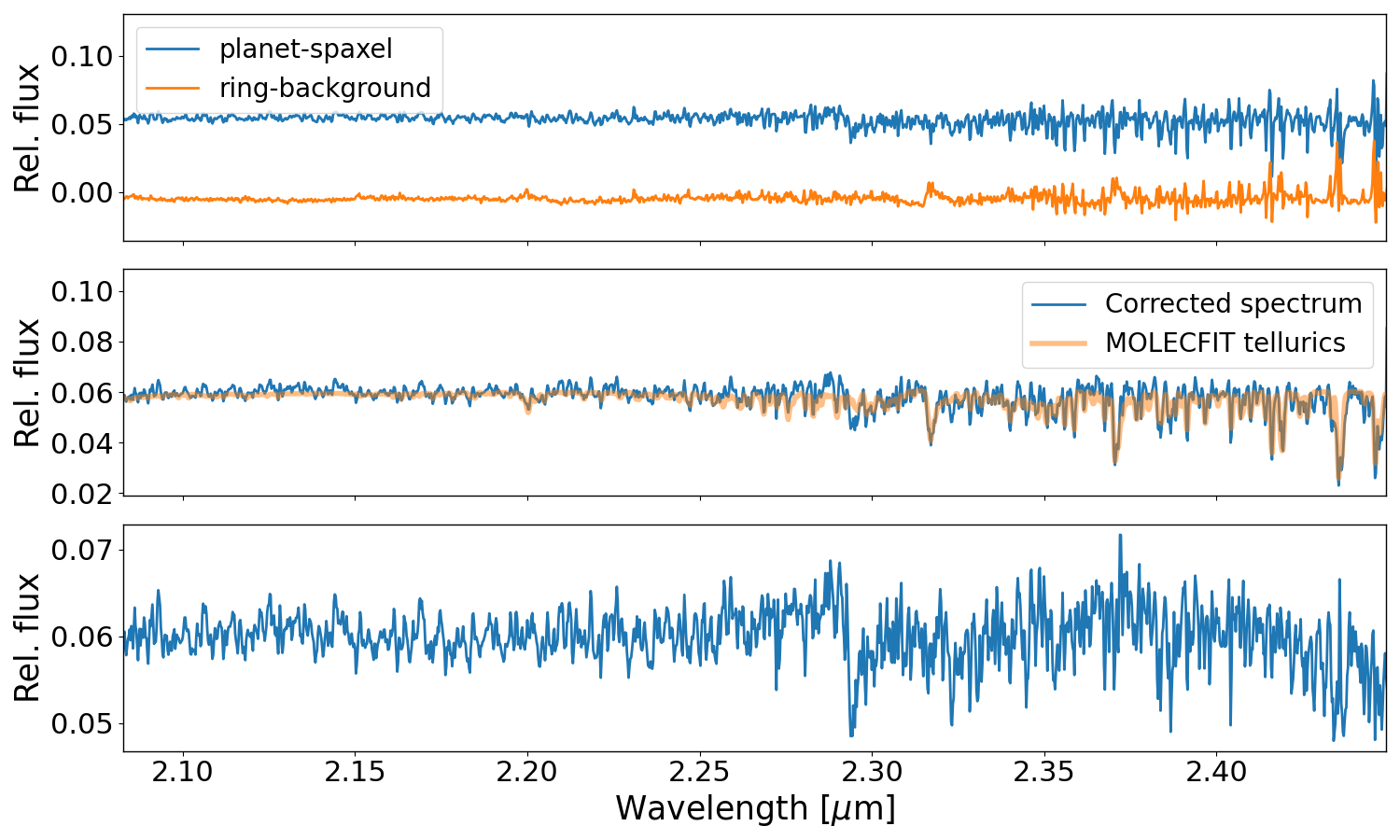}
    \caption{Background subtraction and telluric removal from a planet-spaxel spectrum. Top: planet spaxel spectrum and average background on a distant ring; middle: $\eta_\text{res,corr}$ compared to its fitted telluric model $\eta_\text{tell}$; bottom: planet-spaxel spectrum corrected from background and tellurics.}
    \label{fig:tell_back}
\end{figure}

At each spaxel $i$ of the residual cube, we fit out this telluric model $\eta_\text{model}$ from $\eta_{i,\text{res}}$. To do so, we first need to correct for a wavelength-dependent background contribution whose non-zero level is causing an artificial decrease of telluric lines prominence and the appearance of spurious features, such as telluric lines residuals. Those can be seen by comparing a planet-spaxel spectrum and its surrounding background (Fig.~\ref{fig:tell_back}). The origin of the continuum level of the background is unknown but it could be due to scattered stray light. The features are generated by the differences between the stellar (plus telluric) spectrum actually observed at the given spaxel and the reference stellar (plus telluric) spectrum that was removed in Section~\ref{sec:starspec}\footnote{Spaxel-to-spaxel variations in the telluric spectrum arise from small differences in the wavelength calibration from one spaxel to the others, and from adding a non-zero spaxel-dependent background continuum when fitting and subtracting the reference stellar (plus telluric) spectrum in Section~\ref{sec:starem}}.

We make an estimation of the background contribution in a spaxel $i$ by taking the median spectrum in a distant ring around this spaxel $\eta_{i,\text{ring}}$. Since the SINFONI PSF has FWHM$\sim$4\,spaxels  in radius, we use a ring radius of 6 spaxels with a width of 1 spaxel. This background estimate is then removed from the spaxel $i$ spectrum $\eta_{i,\text{res,corr}}=\eta_{i,\text{res}}-\eta_{i,\text{ring}}$. Its contribution level at $\sim$2.165\,$\mu$m can be viewed in Fig.~\ref{fig:fov_star_contrib} on the brighter areas away from the planet location. Fig.~\ref{fig:hist_flux} also shows the 2.165-$\mu$m absolute background flux that is $\sim$1-10\% of the absolute total flux at every spaxel. Around the planet location its continuum level is very close to zero, but non-negligible features are seen that have corresponding features in the spectrum at the planet central spaxel. 

Adding a background term to eq.~\ref{eq:plan_spax}, the background subtraction can be summarized with the following set of equations:

\begin{align}
\eta_{i,\text{res,corr}} = &\left(K_{i,p,2.165} + K_{i,b,2.165} - K_{i,p} - K_{i,b} \right) \eta_\star + K_{i,p}\eta_p + K_{i,b}\eta_b \nonumber \\
& -\left(K_{\text{ring},b,2.165}-K_{\text{ring},b}\right) \eta_\star - K_{\text{ring},b}\eta_b
\end{align}

Assuming an equal contribution of background in the central and the neighboring spaxels, i.e. $K_{i,b}=K_{\text{ring},b}$, this equation simplifies to:

\begin{align}
\eta_{i,\text{res,corr}} = & \left(K_{i,p,2.165} - K_{i,p} \right) \eta_\star + K_{i,p}\eta_p
\end{align}

We recover Eq.~\ref{eq:plan_spax}, and the previously implicit background is now explicitly suppressed. 

The final step is to fit $\eta_\text{model}$ the telluric spectrum model to $\eta_{i,\text{res,corr}}$. We use a simple least-squares optimisation at each spaxel, allowing the telluric model to vary in intensity with $\eta_{i,\text{tell}}=a\, \eta_\text{model} + b$ with $a+b=c_i$ if $c_i$ is the level of the residuals at spaxel $i$. Because $\eta_\text{model}$ was determined from the reference stellar spectrum that could be a little affected by background, an offset $b$ is added to account for small intensity differences in telluric lines between $\eta_\text{model}$ and the telluric spectrum at spaxel $i$. We divide out $\eta_{i,\text{tell}}$ from $\eta_{i,\text{res,corr}}$ leading to the telluric-free spectrum $\eta_{i,\text{res, tell-free}}$. 

A summary of background subtraction and telluric removal at the central planet spaxel is shown in Fig.~\ref{fig:tell_back}.

\section{Molecular mapping dedicated to SINFONI spectra}
\label{sec:molmap}

\subsection{Exo-REM grid of models}
\label{sec:exorem}
To calculate the molecular mapping of diverse species in the atmosphere of $\beta$\,Pic\,b, as is now most usually performed (see e.g. ~\citealt{Snellen2014, Brogi2018, Hoeijmakers2018, Petit2018, Ruffio2019, Cugno2021, Petrus2021, Patapis2022, Malin2023}), we calculate a CCF of the observed spectrum at each spaxel with a synthetic spectrum. Here, the templates are taken from an Exo-REM model grid ~\citep{Baudino2015,Charnay2018,Blain2021}. Exo-REM is a 1D radiative-convective model, which self-consistently computes the thermal structure, the atmospheric composition, the cloud distribution and spectra. The model includes the opacities of H$_2$O, CH$_4$, CO, CO$_2$, NH$_3$, PH$_3$, TiO, VO, FeH, K and Na, and collision-induced absorptions of H$_2$–H$_2$ and H$_2$–He. Silicate and iron clouds are included using simple microphysics to determine particle sizes~\citep{Charnay2018}. The model grid includes four free parameters totaling 9573 spectra with T$_\text{eq}$ ranging from 400 to 2000 K with steps of 50 K; $\log g$ from 3.0 to 5.0 dex with steps of 0.5 dex; [M/H] from -0.5 to 1.0 dex with steps of 0.5 and [C/O] from 0.1 to 0.8 with steps of 0.05. The synthetic spectra were computed at $R$=20,000, a compromise between reducing computation speed and reaching the highest resolution possible on atmospheric models to be used as templates of low-to-medium resolution spectra of exoplanets.

\begin{figure}
    \centering
    \includegraphics[width=89.3mm]{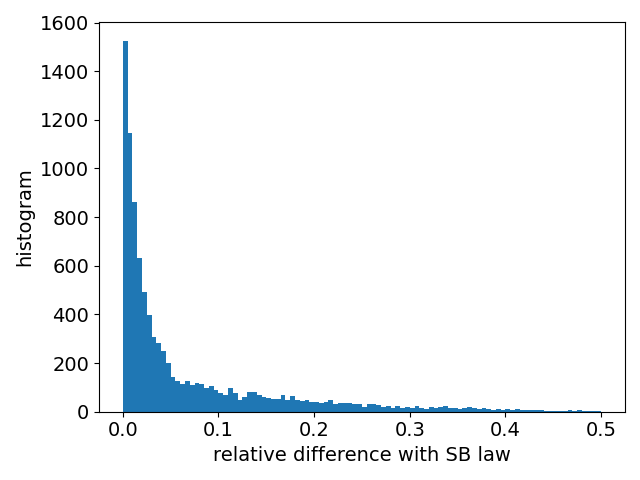}
    \caption{Histogram of relative difference between the exo-REM models emittance and Stefan-Boltzman law.}
    \label{fig:hist_exorem}
\end{figure}

We cleansed the model grid from those that did not converge well. To do so, we calculated for each spectrum the integral on the whole wavelength domain from visible to far IR, $I_s$. We compared this integral to the theoretical $\sigma \,T_\text{eff}^4$ Stefan-Boltzman law. This deviation peaks below $\sim$5\% (Fig.~\ref{fig:hist_exorem}). We thus removed 3315 among 9573 spectra (i.e. 35\%) with a deviation larger than 5\%. 

Each synthetic spectrum produced by Exo-REM results from individual contributions (or spectrum) of the species out of which it is composed, mainly CO, H$_2$O, CH$_4$ and NH$_3$, which we can use as individual molecular templates.

\subsection{The cross-correlation function of the planet spectrum}
\label{sec:CCF}
\begin{figure*}
    \centering
    \includegraphics[width=178.6mm]{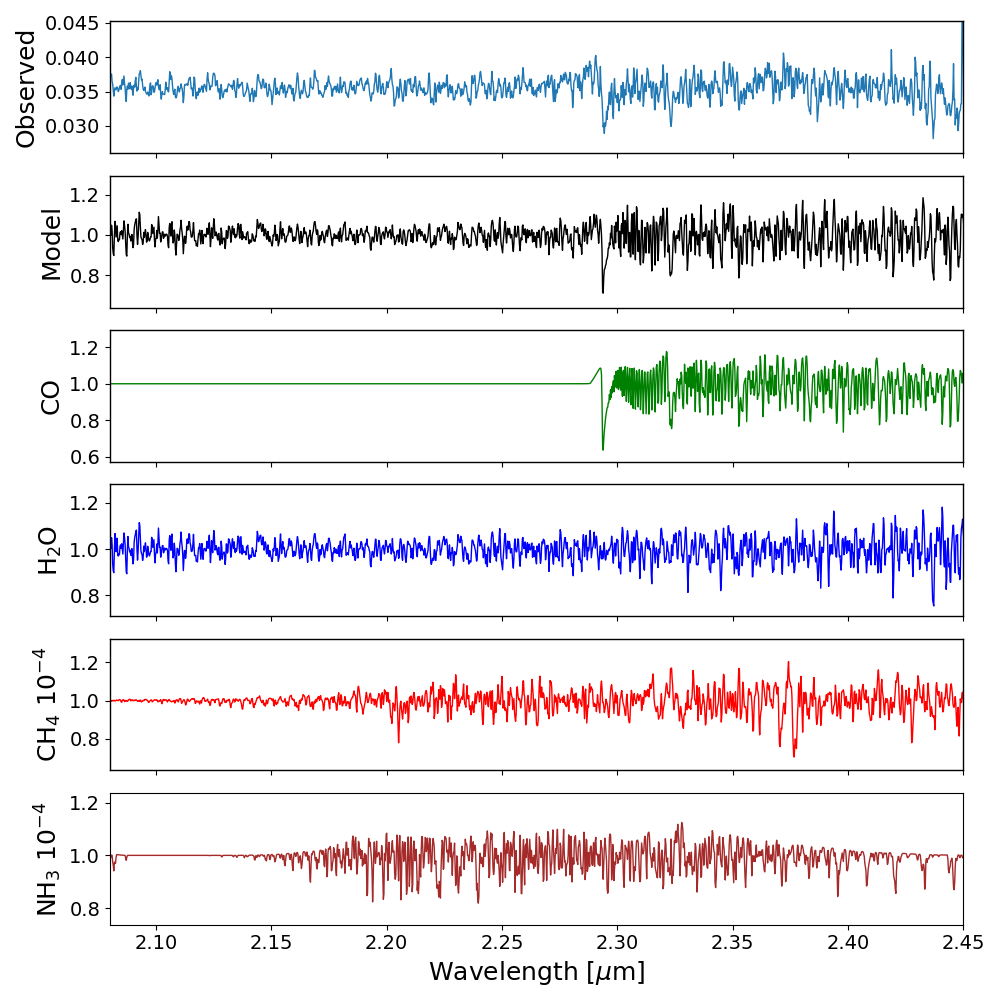}
    \caption{Top: the SINFONI planet absorption spectrum of $\beta$\,Pic\,b along the K-band once corrected from star, background and telluric pollutions. Bottom: Exo-REM spectrum of simulated atmosphere of 1700\,K, $\log(g)$=4.0 and $[$Fe/H$]$=0.0 with $R$=6000. Individual contributions of species CO, H$_2$O, CH$_4$ and NH$_3$ are represented from top to bottom with diverse colors. For all the spectra, their continuum were divided out as explained in Section~\ref{sec:CCF}.}
    \label{fig:planet_spectrum}
\end{figure*}

At every spaxel, between 2.08 and 2.43\,$\mu$m, we calculate the CCF of the observed spectrum and an Exo-REM synthetic spectrum for an assumed $T_{\text{eff}}$=1700\,K, $\log g$=4.0 cgs and $[$Fe/H$]$=0.0 dex, as based on Table 3 in~\citet{Nowak2020}. The expected abundances of NH$_3$ and CH$_4$ in the atmosphere $\sim$\,\!$10^{-6}$ are too low to yield absorption features detectable in the SINFONI spectra. We thus artificially enhance their abundances to 10$^{-4}$, in order to probe possible over-abundance of these species in $\beta$\,Pic\,b's atmosphere. Details on the contributions of H$_2$O, CO, NH$_3$ and CH$_4$ are shown in Fig.~\ref{fig:planet_spectrum} and compared to the $\beta$\,Pic\,b's spectrum derived in next Section~\ref{sec:average}.

We exclude the red edge of the K-band beyond 2.43\,$\mu$m which displays the strongest telluric lines remnants. Prior to the calculation, we divide the continuum of the observed and synthetic spectra. The continuum is obtained by first fitting a 4th-degree polynomial, and then applying a median filter with a window-width of 0.01\,\AA, combined to a smoothing Savitzky-Golay filter of order 1. The CCF is calculated by directly cross-correlating the median-removed observed and synthetic spectra at different shifts. The CCF is finally normalised by the norm of the spectra, thus leading to a zero-normalised CCF. This results in the molecular maps shown in Fig.~\ref{fig:molmap}.

\setlength{\unitlength}{1mm}
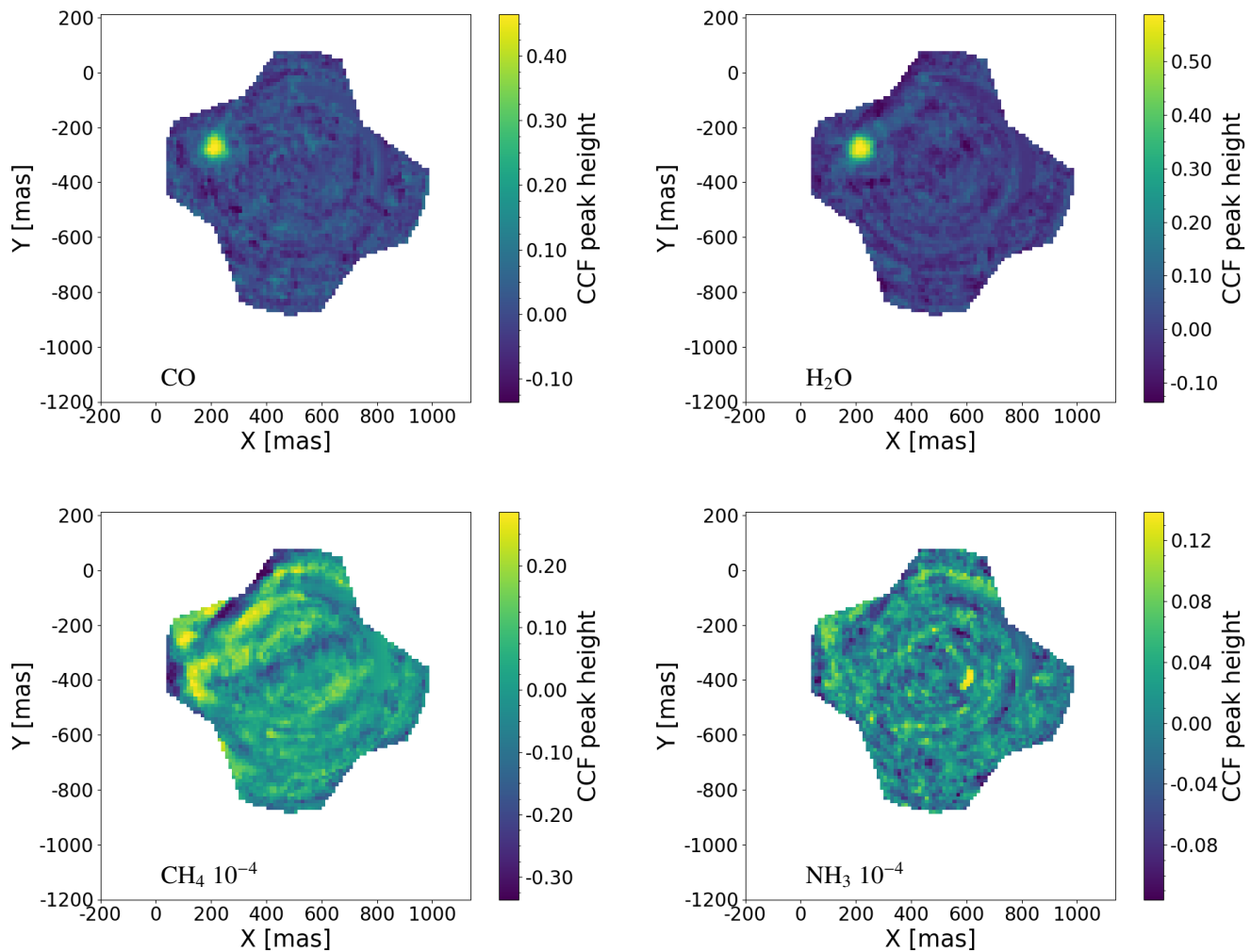
\begin{figure*}
    \begin{picture}(188.6,140)
    \put(0,70){\includegraphics[width=89.3mm,clip=true,trim=70 0 40 60]{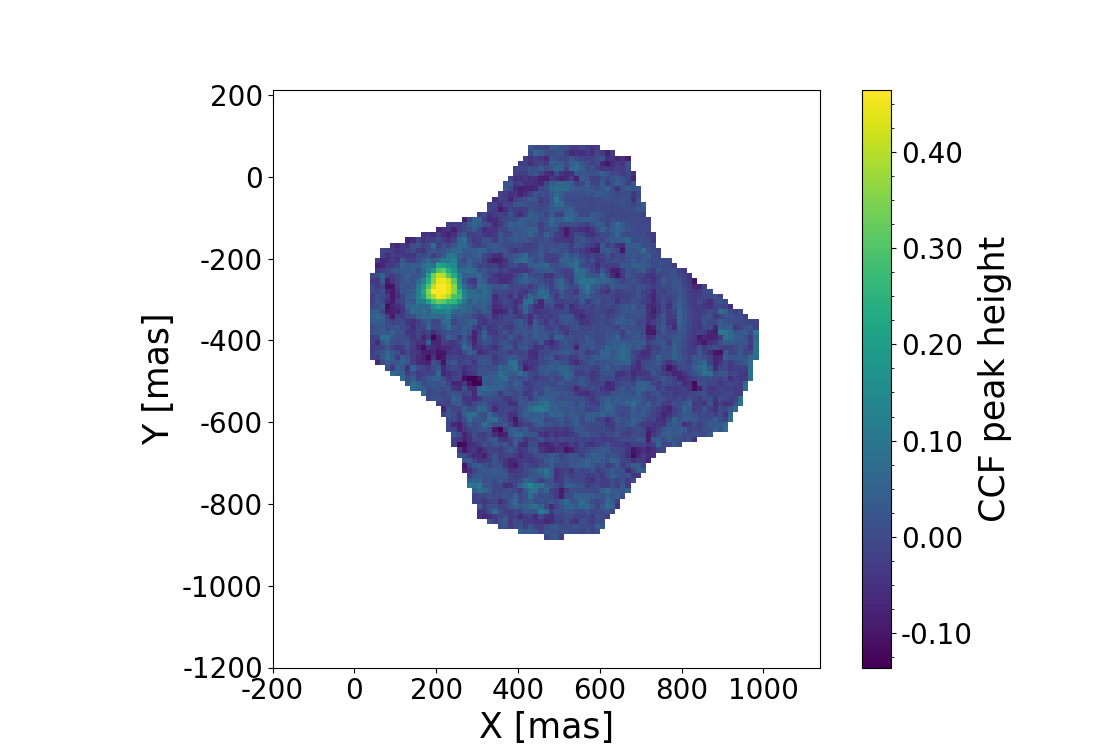}}
    \put(25,80){CO}
    \put(90,70){\includegraphics[width=89.3mm,clip=true,trim=70 0 40 60]{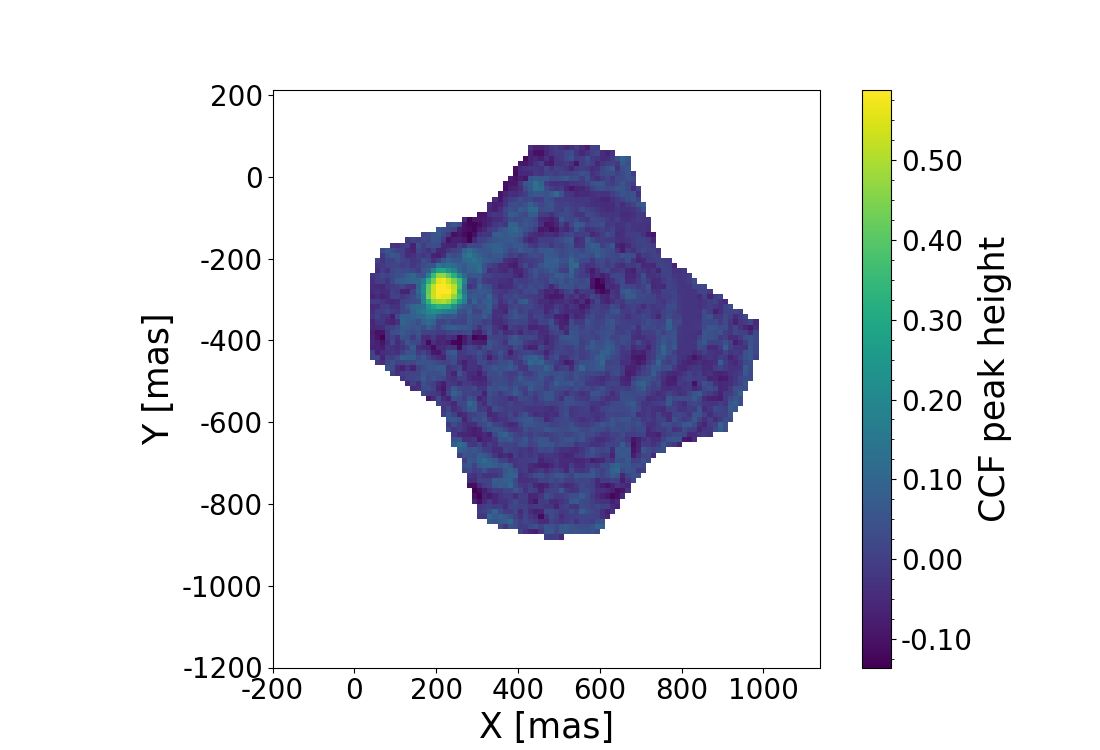}}
    \put(115,80){H$_2$O}
    \put(0,0){\includegraphics[width=89.3mm,clip=true,trim=70 0 40 60]{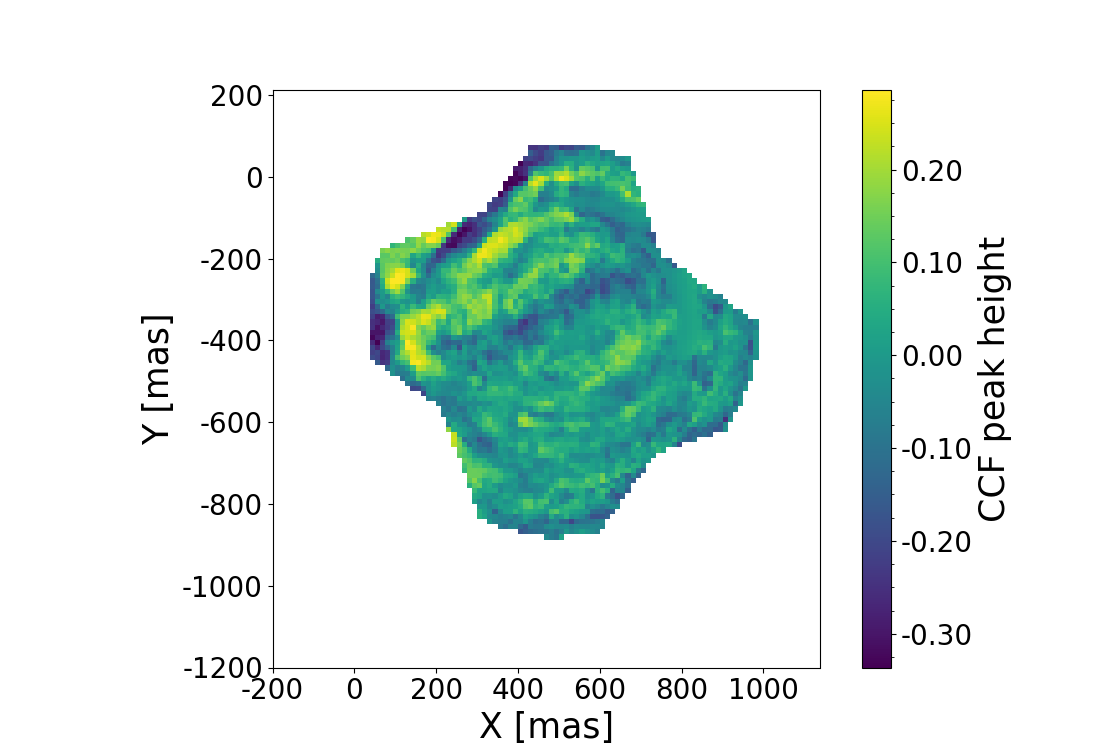}}
    \put(25,10){CH$_4$ $10^{-4}$}
    \put(90,0){\includegraphics[width=89.3mm,clip=true,trim=70 0 40 60]{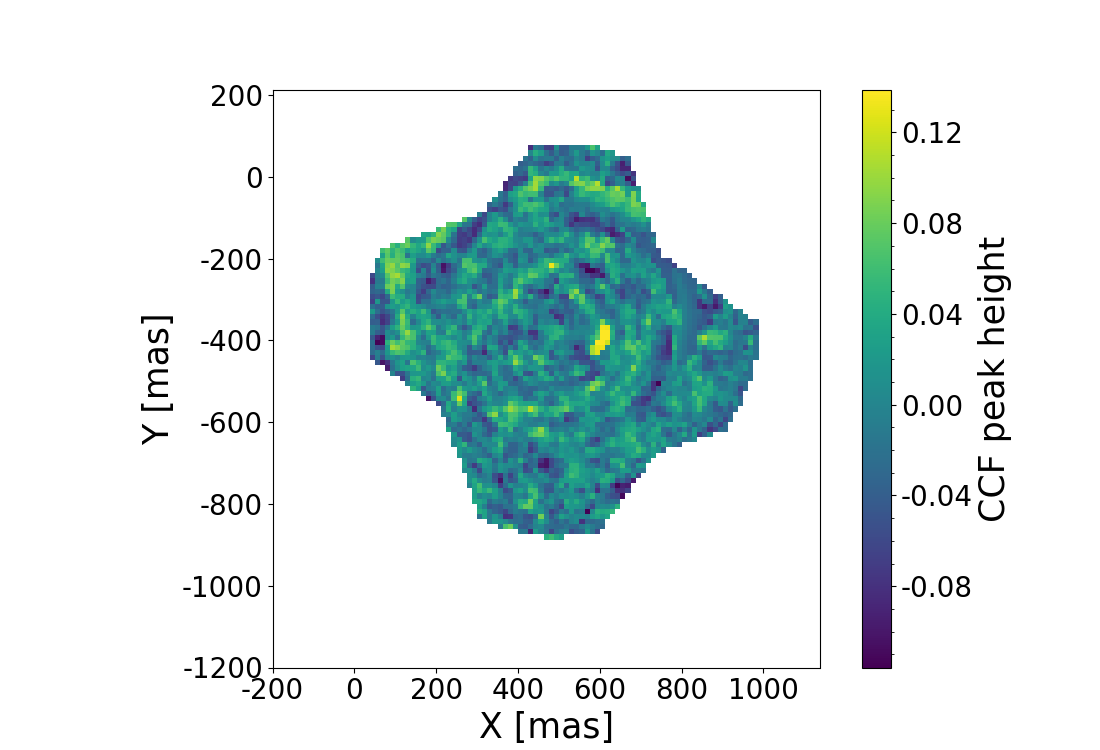}}
    \put(115,10){NH$_3$ $10^{-4}$}
    \end{picture}
    \caption{Molecular CCF maps of $\beta$\,Pic\,b for species CO, H$_2$O, CH$_4$ (10$^{-4}$) and NH$_3$ (10$^{-4}$). At each spaxel, the maps show the CCF height at 0\,km\,s$^{-1}$, a value close to the expected radial velocity of the planet $b$, $\left|\text{VR}\right|$$<$10\,km\,s$^{-1}$given the instrument LSF of $\sim$75\,km\,s$^{-1}$. }
    \label{fig:molmap}
\end{figure*}

We then fit the PSF of the planet CCF halo by a 3D Gaussian with respect to $(\delta,\alpha,v_r)$. This determines the position and radial velocity of the planet, as well as the broadening of the PSF and of the LSF. The results are summarized in Table~\ref{tab:SN_table}. They are compared to the values derived in the same fashion but following H18's recipe to suppress the stellar pollution. We only show results for CO, H$_2$O and using the full spectrum, confirming, as in H18, the presence of CO and H$_2$O but not detecting NH$_3$ and CH$_4$. With some variability from one model to the other, the CCF peak is located at a separation to the central star of $\sim$351$\pm$5\,mas, with a PSF broadening standard deviation of $\sim$2.7\,spaxels, that is 34\,mas.

A signal-to-noise (S/N) ratio is calculated as the height of the peak divided by the fit residuals on a volume of 200 pix$^2$ by 2000\,km\,s$^{-1}$ around the planet ($\delta$,$\alpha$,$v_r$)-location. Our \verb+starem+ method leads to S/N comparable to those obtained through H18 method. We did not use SYSREM as they did to remove spaxel-to-spaxel correlated noise within the cubes, but corrected for the background differently as explained in Section~\ref{sec:back}. We derived slightly better S/N by subtracting the star halo using STAREM than those obtained using the H18 subtraction method, leading to S/N$_\text{all}$=19.6, S/N$_\text{CO}$=11.8, S/N$_\text{H$_2$O}$=15.4. This follows the trend found also for HIP\,65426\,b~\citep{Petrus2021} where taking into account all species leads to the best detection S/N, while for individual contributions of species, H$_2$O gives significantly better results than CO. This last properties is best explained by the presence of less prominent but more numerous lines of H$_2$O compared to CO all along the K-band, and especially away from regions spanned by telluric lines.

We note that in the H18's paper, the S/N they obtained are larger (all molecules: 22.8 vs. 17.5 here; CO: 13.7 vs. 11.2 here; H$_2$O: 16.4 vs. 15.7 here). This difference can be explained, because they averaged the CCF of cubes obtained at two different nights, while we used the cubes from only a single night. The smaller difference of S/N for H$_2$O can be explained by the presence of traces of telluric lines in the H18 residual cube because they did not recalibrated the wavelength solution before subtracting the reference spectrum (containing the tellurics). We also note that the calculation performed in H18's paper to determine the S/N is slightly different in that they use a distant 3D-ring around the CCF peak to estimate the noise, while we calculate the noise from the residuals of a fit of the CCF peak.

\begin{table*}[hbt]\centering
    \caption{\label{tab:SN_table} Comparing results of the different star spectrum subtraction schemes on the normalised CCF of the residual map with a 1700\,K, $\log g$=4.0, $[$Fe/H$]$=0.0 Exo-REM model, detailing the CO and H$_2$O individual contributions. We also tried the molecular mapping on CH$_4$, NH$_3$, FeH, and CO$_2$, none leading to a significant detection within the area around planet spaxels. 
    }
    \vspace{0.5cm}
    \begin{tabular}{lccc|cc|ccc}
        Method & \multicolumn{3}{c|}{CCF strength} & \multicolumn{2}{c}{Radial velocity} & \multicolumn{3}{c}{Planet PSF}\\ 
         & peak & noise & S/N & $v_r$     & $\sigma_{\text{broad}}$   & $X_c$\tablefootmark{$\dagger$}  & $Y_c$\tablefootmark{$\dagger$} & $\sigma_{\text{PSF}}$ \\
         & height & level &     & (km/s) & (km/s) & (mas) & (mas) & (pixel)  \\
        \hline
        \multicolumn{9}{c}{Full-model: 1700\,K, $\log g$=4.0, $[$Fe/H$]$=0.0, $R_{\text{spec}}$=6000} \\ \\
        H18 & 0.8952$\pm$0.0076 & 0.0511 & 17.52$\pm$0.15 & 1.03$\pm$0.35 & 46.80$\pm$0.38 & 218.0$\pm$0.25 & -277.13$\pm$0.25 & 2.34$\pm$0.01\\
        STAREM & 0.8452$\pm$0.0060 & 0.0431 & 19.60$\pm$0.14 & -0.25$\pm$0.25 & 41.87$\pm$0.30 & 216.63$\pm$0.25 & -276.75$\pm$0.25 & 2.71$\pm$0.01 \\
        \hline
        \multicolumn{9}{c}{only CO} \\ \\
        H18 & 0.4837$\pm$0.0049 &  0.0432 & 11.20$\pm$0.11 & -4.29$\pm$0.41 & 45.91$\pm$0.49 & 215.38$\pm$0.25 & -274.25$\pm$0.25 & 2.47$\pm$0.02  \\
        STAREM & 0.5094$\pm$0.0048 & 0.0431 & 11.81$\pm$0.11 & -3.24$\pm$0.35 & 41.85$\pm$0.42 & 214.00$\pm$0.25 & -273.50$\pm$0.25 & 2.66$\pm$0.02 \\
        \hline
        \multicolumn{9}{c}{only H$_2$O} \\ \\
        H18 & 0.7887$\pm$0.0076 & 0.0523 & 15.07$\pm$0.15 & 2.28$\pm$0.38& 41.64$\pm$0.33 & 218.62$\pm$0.25 & -278.63$\pm$0.25 & 2.28$\pm$0.02 \\
        STAREM & 0.6947$\pm$0.0056 & 0.0451 & 15.41$\pm$0.12 & 0.76$\pm$0.29 & 41.64$\pm$0.33 & 217.13$\pm$0.25 &   -279.00$\pm$0.25 & 2.75$\pm$0.02\\
        \hline
    \end{tabular}
    \tablefoot{\\
    \tablefoottext{$^\dagger$}{The star is located at coords $(X_\star,Y_\star)$=(0,0) mas.}
    }
\end{table*}

\section{Extracting the planet atmosphere absorption spectrum}
\label{sec:average}

\begin{figure}
    \centering
    \includegraphics[width=89.3mm]{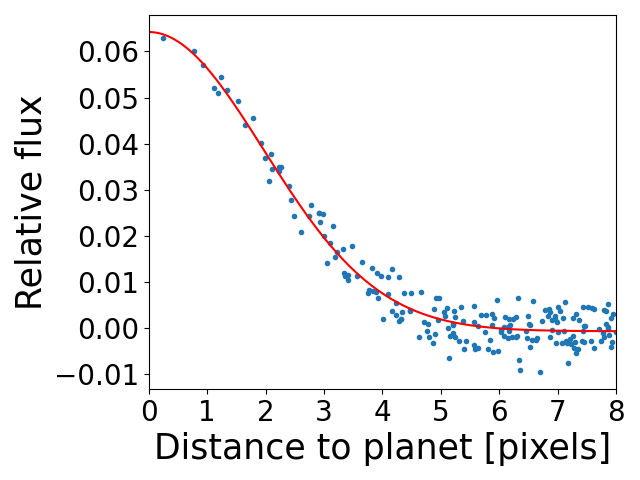}
    \caption{The planet PSF radial profile at a varying distance from 0 to 8 spaxels. The relative flux is obtained by dividing out the continuum and subtracting the star spectrum, as explained in Section~\ref{sec:starem}.}
    \label{fig:planet_profile}
\end{figure}

The PSF of the planet follows an Airy profile whose central region can be approximated by a Gaussian profile. This can be seen in Fig.~\ref{fig:planet_profile} where the residual flux of the stellar subtraction at 2.165\,$\mu$m is shown with respect to the distance of the spaxels to the planet center position. The planet center position is obtained by fitting the residual flux at 2.165\,$\mu$m by a 2D-Gaussian. The radial profile of the planet PSF is well modeled by a 2-spaxel wide Gaussian. The planet relative flux rises up to 7\% near the center but becomes negligible compared to noise beyond a distance of $\sim$5 spaxels. 

The average spectrum at those spaxels is calculated by integrating the relative flux on all spaxels up to a given radius and applying to each spaxel a weight that depends on its distance to the planet. We model the PSF profile by the Gaussian used above, with $F(r)$=$e^{-r^2/2\sigma_\text{PSF}^2}$ where $r$=$\Vert{\bf r_i}-{\bf r_c}\Vert$ is the spaxel-distance between a given spaxel $i$ and the planet center position, and $\sigma_\text{PSF}$=2\,spaxels at 2.165\,$\mu$m. We use this profile as the weight function with thus $w(r)$=$e^{-r^2/2\sigma_\text{PSF}^2}$. When integrating the spaxels flux, both planet and noise signals grow, yet limited by the PSF flux dimming, and the signal-to-noise increases like $\sqrt{N_\text{spaxels}}$. We found reasonable to integrate the flux up to 4--$\sigma$ from the PSF centroid, with $\sigma(\lambda)$=$\sigma_\text{PSF} \times 2.165\,\mu$m/$\lambda$. 

\begin{figure}
    \centering
    \includegraphics[width=89.3mm]{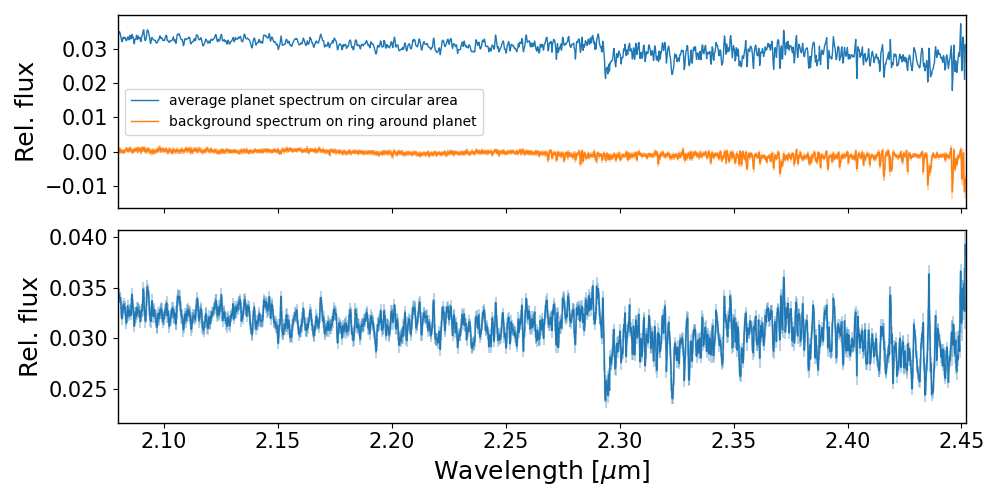}
    \caption{Background removal. Top panel: average spectrum about the planet centroid (blue) compared to the average background spectrum on a ring around the planet (orange). Common features can be well distinguished. Bottom panel: planet spectrum corrected from the background. The features remaining are $\beta$\,Pic\,b's atmosphere absorption lines, while the main stellar feature (Brackett-$\gamma$) has been well removed.}
    \label{fig:background}
\end{figure}

Since we cannot assume that the background removal was absolutely perfect in Section~\ref{sec:back}, we again estimate the background pollution within planet spaxels using neighboring spaxels in a ring around the planet. The background level at distance to the planet centroid is close to zero within $\pm$0.01, as can be seen in the PSF profile, Fig.~\ref{fig:planet_profile}. We use the median spectrum within this ring as an estimation of the background spectrum shown in the top panel of Fig.~\ref{fig:background}. This background spectrum is subtracted from the average planet spectrum, leading to the corrected planetary spectrum, also shown Fig.~\ref{fig:background}. We found the ring radii optimising the S/N of the corrected planet spectrum to be 4.2 -- 5.1 spaxels. 

As an ultimate step, the spectrum was divided by the continuum, estimated, as in Section~\ref{sec:CCF}, by applying a median filter with a window-width of 0.01\AA~combined to a smoothing Savitzky-Golay filter of order 1. 
The final planet spectrum obtained is compared to a theoretical Exo-REM spectrum of a 1700\,K planet with $\log g$=4.0\,cgs and $[$Fe/H$]$=0.0\,dex in Fig.~\ref{fig:planet_spectrum}. 

\section{Template-matching of the planet spectrum}
\label{sec:template-match}

We addressed the problem of finding the best matching model of the planet spectrum in two ways: first by exploring the space of available templates and their match with the planet spectrum using a simple grid search, second by running a Markov-Chain Monte Carlo (MCMC) sampling around the optimum spectrum. In both cases we used the forward-modeled Exo-REM~\citep{Charnay2018} spectra models of exoplanet atmosphere.  

\subsection{Grid search}
\label{sec:gridsearch}

We first explored the template space by a grid search. We tested the $\chi^2$ and zero-normalised CCF scores for optimizing the models. Each model spectrum is convolved by a Gaussian Kernel corresponding to a resolving power of $4020$ and by a rotational profile with  $v\sin i$=25\,km\,s$^{-1}$. The rotational profile is the usual bell-like profile with limb-darkening coefficient $\varepsilon$=0.6 (Gray 1997). Each model spectrum is then flattened with the same median  filtering function, with a window width of 0.01\,$\mu$m, as used to flatten the observed spectrum. We moreover corrected for the "warp" effect noted in Section~\ref{sec:starem} in the models by adding a supplementary term in $(1-C_p/C)\,\eta_\star$, with $\eta_\star$ the normalised star spectrum, $C_p$ the model's continuum and $C$ the average continuum in the SINFONI cube at the planet spaxels both normalised to 1 at 2.165\,$\mu$m. For the ZNCCF the median of both model and data is subtracted before cross-correlation. For the $\chi^2$, the fit function includes two other parameters applied to the model spectrum, a scale $a$ to allow compensate for the arbitrary level of the spectrum continuum, and a rigid Doppler shift $v_r$. The fit is then performed using the \verb+curve_fit+ procedure from the \verb+scipy+ library. 

The error bars of data points are derived from the relation $\sigma_{\text{data}}$$\propto$$\sqrt{F_{\star,\text{approx}}}/C$, with $F_{\star,\text{approx}}(\lambda)$ the reference stellar spectrum defined in Section~\ref{sec:starspec} and $C(\lambda)$ the continuum by which spectra are divided in Section~\ref{sec:starem}. The level of noise is normalised to the error directly measured in the planetary spectrum at 2.165\,$\mu$m from the standard deviation of flux to the median on a width of 0.003\,$\mu$m ; we take the average of the errors at all spectral channels from 2.15 to 2.18\,$\mu$m.  

\begin{figure}
    \centering
    \includegraphics[width=89.3mm, clip=true, trim=10mm 10mm 30mm 30mm]{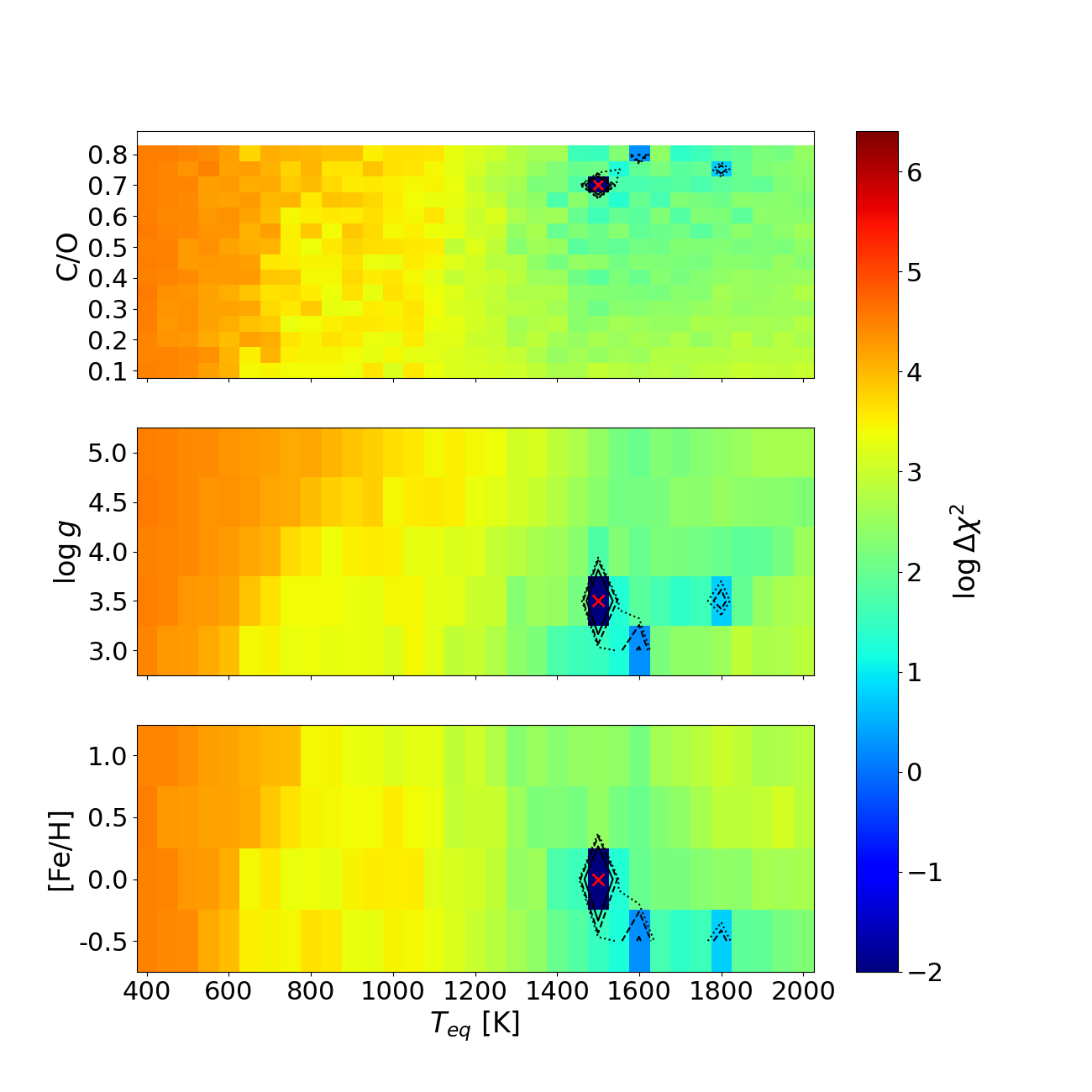}
    \caption{Grid search $\Delta\chi^2$ maps for T$_{\text{eff}}$-$\log g$ (top panel) T$_{\text{eff}}$-C/O (middle panel) and T$_{\text{eff}}$-$]$Fe/H$]$. The solid, dashed and dotted lines indicate 1, 2 and 3-$\sigma$ confidence regions. The red cross locates the optimum.}
    \label{fig:chi2maps}
\end{figure}

\begin{figure}
    \centering
    \includegraphics[width=89.3mm, clip=true, trim=10mm 10mm 30mm 30mm]{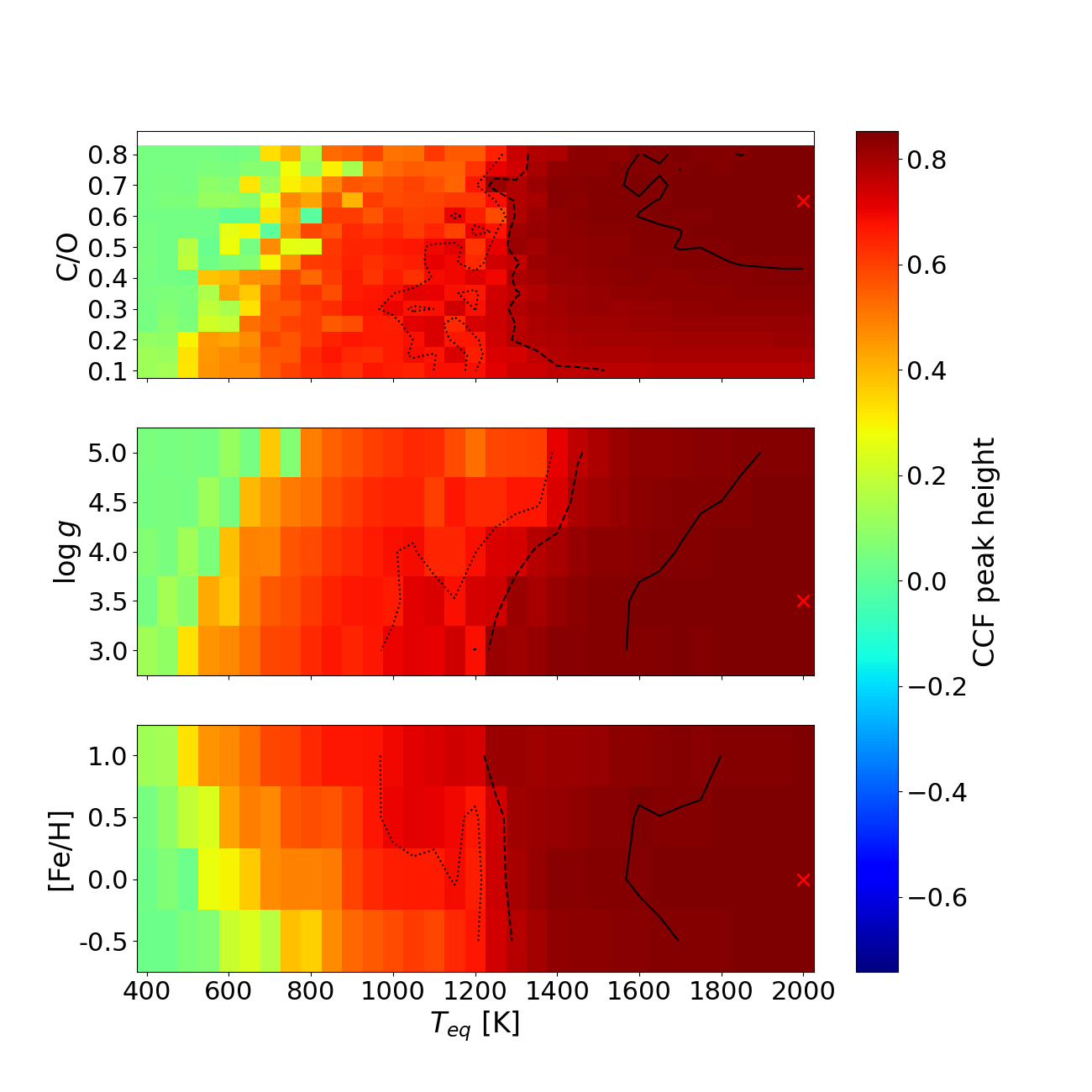}
    \caption{Grid search CCF maps for T$_{\text{eff}}$-$\log g$ (top panel) T$_{\text{eff}}$-C/O (middle panel) and T$_{\text{eff}}$-$]$Fe/H$]$. The solid, dashed and dotted lines indicate 1, 10 and 20\% $\Delta$CCF levels compared to the optimal CCF. The red cross locates the optimum.}
    \label{fig:ccfmaps}
\end{figure}

The grid search maps with $\chi^2$-score and CCF-score are shown on respectively Fig.~\ref{fig:chi2maps} and~\ref{fig:ccfmaps} and the results are summarized in Table~\ref{tab:chi2gridsearch}. For the $\chi^2$ the confidence intervals are bounded by the $\Delta\chi^2$ with the 1, 2 \& 3--$\sigma$ regions corresponding to $\Delta\chi^2$=2.3, 6.2 \& 11.8. For individual measurements in Table~\ref{tab:chi2gridsearch} the 1 and 2--$\sigma$ intervals correspond to $\Delta\chi^2$=1 \& 4, interpolated from the 1D $\Delta \chi^2$ obtained for each parameter. The small extent of the 1-$\sigma$ confidence regions around the best fit model is most likely an effect of the discrete grid used to explore the parameter space. Half the 2-$\sigma$ intervals are certainly more reliable than the 1-$\sigma$ intervals as error bars. For the CCF grid search we show the $\Delta$CCF regions of 1, 10 and 20\% ; they are not translated into confidence intervals and only the optimum model is given in Table~\ref{tab:chi2gridsearch}.

\begin{figure}
    \centering
    \includegraphics[width=89.3mm,clip=true,trim=0 0 0 22pt]{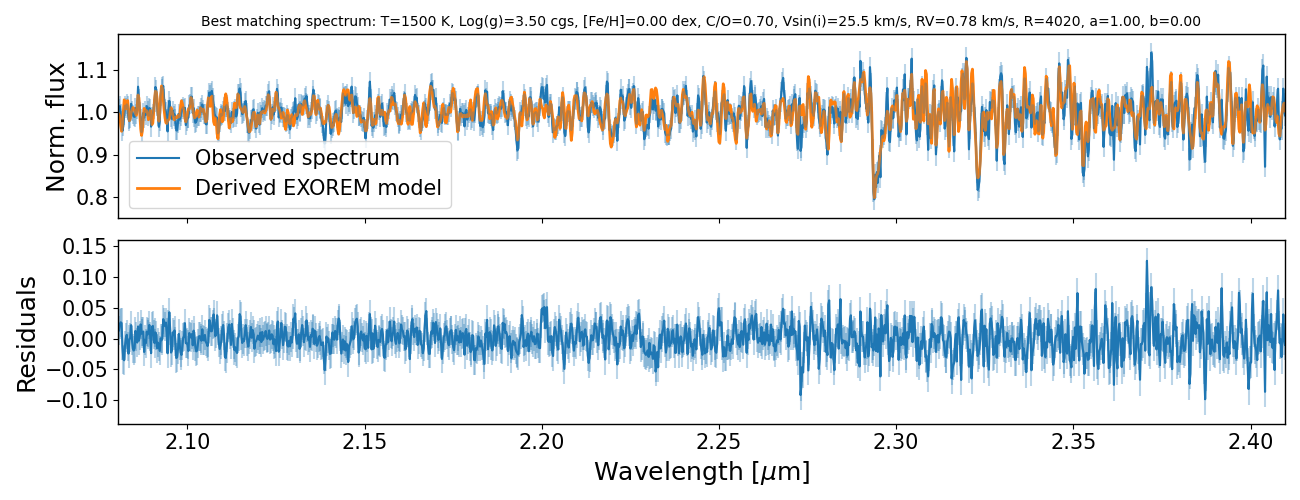}
    \includegraphics[width=88.mm,clip=true,trim=0 0 10pt 22pt]{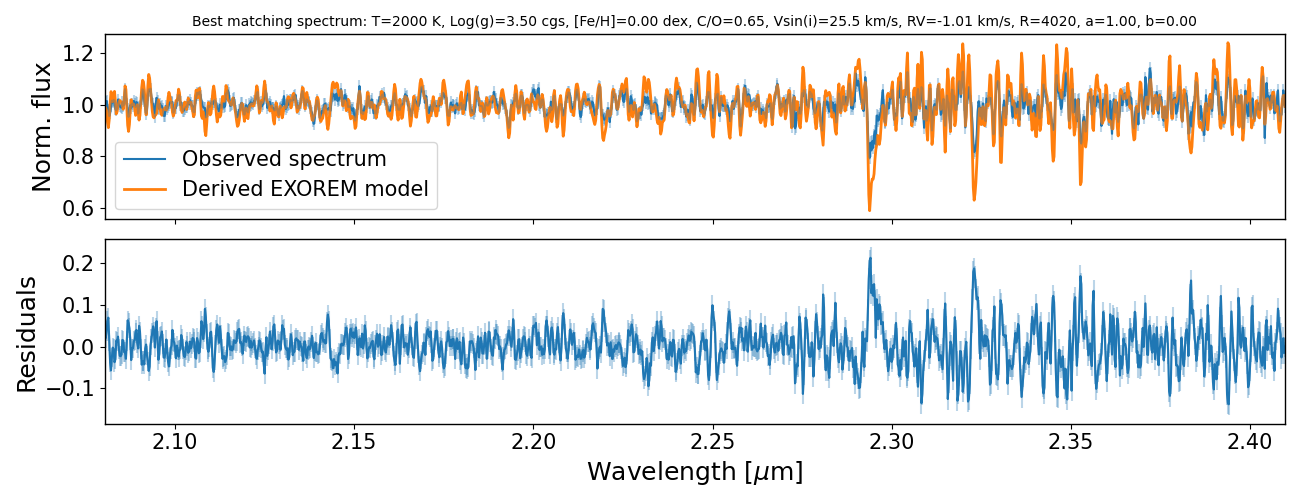}
    \caption{Best matching models, using the $\chi^2$ optimisation (top) and the CCF optimisation (bottom).}
    \label{fig:bestgridspec}
\end{figure}

The $\chi^2$ grid search leads to a solution with $T_{\text{eff}}$=1550\,K, $[$Fe/H$]$=0.0\,dex, $\log(g)$$\sim$$3.5$, and C/O$\sim$0.70. The best matching model is compared to the observed spectrum in Fig.~\ref{fig:bestgridspec}. The CCF grid search leads to a different solution with much larger temperature saturating at 2000\,K and a solar metallicity. But in this case, as shown in Fig.~\ref{fig:bestgridspec} the spectra do not match in line amplitude. This shows that the CCF, due to the removal of the continuum and the normalisation of the spectra, is not well adapted to grid search with a strong degeneracy of models.

\begin{table*}
    \centering
    \caption{$\chi^2$ and CCF grid search results (see Section~\ref{sec:gridsearch}. The reduced optimum $\chi^2$ has been normalised to 1 and is therefore not given here. Uncertainties are obtained by interpolation of the 1D $\Delta\chi^2$ maps.}
    \begin{tabular}{lc|ccc|c}
         Parameter & Unit & \multicolumn{3}{c|}{$\chi^2$ grid} &  CCF grid \\
         & & optimum & 1-$\sigma$ &  2-$\sigma$ & optimum \\
         \hline
         $T_{\text{eff}}$ &  (K) & 1500 & (+3)(-1) &  (+100)(-5) & 2000 \\
         $\log g$ & (cgs) & 3.5 & (+0.0)(-0.3) & (+0.0)(-0.5) & 3.5 \\
         $[$Fe/H$]$ & (dex) &  0.0 & (+0.0)(-0.3) & (+0.0)(-0.5) & 0.0 \\
         C/O & & 0.70 & (+0.01)(-0.00) & (+0.10)(-0.01) & 0.65 \\
         \tablefootmark{$\dagger$}$v_r$ & (km\,s$^{-1}$) & 0.78 & $\pm$0.79 & &  -1.10  \\
         \tablefootmark{$\dagger$}$a$ & & 0.999 & $\pm$0.025 & & fixed to 1 \\
         \tablefootmark{$\dagger$}$v\sin i$ & (km\,s$^{-1}$) &  fixed to 25 &   & & fixed to 25  \\
         R & & fixed to 4020 & & & fixed to 4020 \\
         \hline
         $\chi^2_r$ & & 1 & &\\
         max(CCF) & & & & & 0.69 \\
         \hline
    \end{tabular}
    \label{tab:chi2gridsearch}
    \tablefoot{\\
    \tablefoottext{$\dagger$}{The 1-$\sigma$ error bars for $v_r$, $v\sin i$, and $a$ are obtained for the optimal model using scipy's \texttt{curve\_fit} procedure. The $v\sin i$ and $a$ are not optimized for the CCF, assuming $v\sin i$=25\,km\,s$^{-1}$ and $a$=1.}
    }
\end{table*}

\subsection{An MCMC sampling of models}
\label{sec:mcmc}
We use \verb+emcee+~\citep{Foreman2013} to apply 1,000,000 iterations of Markov-Chain Monte-Carlo (MCMC) sampling around the best solution found by gridsearch in Section~\ref{sec:gridsearch} to assess the true posterior distribution and correlations among the varied parameters. The fitted physical parameters are $T_\text{eff}$, $\log g$, $[$Fe/H$]$, C/O, $v_r$ and $v\sin i$. We also fit for the spectral resolving power $R_\text{spec}$, in order to account for possible deviations from $R$=4020$\pm$30. We also sample a rescaling $a$ to adjust the continuum of model and observed spectra together. The pre-processings of data and models were the same as those used adopted in the gridsearch. The error bars of the data points $\sigma_\text{data}$ (as defined in Section~\ref{sec:gridsearch}) might be over or under estimated by a certain multiplicative amount $f_\text{err}$ and under-estimated by a constant jitter $\sigma$. We thus use the error model $\sigma_\text{err}^2=(f_\text{err} \sigma_\text{data})^2 + \sigma^2$ with $\sigma_\text{data}$ depending on wavelength. We add $f_\text{err}$ and $\sigma$ as hyper-parameters in the MCMC. The prior distributions, either normal ${\mathcal N}$ or uniform ${\mathcal U}$, assumed for all parameters are listed in Table~\ref{tab:priors_MCMC}. All parameters, except $R_{\text{spec}}$, follow uniform priors.

\begin{table}
    \centering
    \caption{List of prior probability density functions (PDF) and parameters used for the MCMC runs. For $T_{\text{eff}}$ two types of priors are used, either uniform along the range of possible values. The MCMC run stops whenever the maximum autocorrelation length among all parameters, $\max(\tau_{\lambda})$, is larger than 50 times the actual number of iterations.}
    \label{tab:priors_MCMC}
    \begin{tabular}{c|c}
        Parameter & Prior PDFs or value\\
        \hline 
        $T_{\text{eff}}$ & ${\mathcal U}(400;2000)$ or ${\mathcal N}(1724;15)$\\
        $\log g$ & ${\mathcal U}(3.0;5.0)$ or ${\mathcal N}(4.18;0.01)$\\
        $[$Fe/H$]$ & ${\mathcal U}(-0.5;1.0)$\\
        C/O & ${\mathcal U}(0.1;0.8)$\\
        $v\sin i$ & ${\mathcal U}(0,+\infty)$ \\
        $R_{\text{spec}}$ & ${\mathcal N}(4020;30)$ \\
        $v_r$ & ${\mathcal U}(-\infty,+\infty)$\\
        $f_{\text{err}}$ & ${\mathcal U}(0.1,10)$ \\
        $\sigma$ & ${\mathcal U}(0,0.5)$ \\
        $a$ & ${\mathcal U}(0.5,1.5)$ \\
        \hline \\
        $N_{\text{steps, max}}$ & 1,000,000 \\
        $N_{\text{walkers}}$ & 20 \\
        \hline
    \end{tabular}
\end{table}

The Exo-REM model spectra are interpolated through the initial 4D-grid at the specific values of parameters chosen by the MCMC walkers at each new step. For a quick calculation, we apply the \verb+LinearNDInterpolator+ from \verb+scipy+ on the smallest Hull simplex, where each vertex is a model of the Exo-REM grid, surrounding each MCMC sampled points. This accounts for missing models at several grid nodes (see Section~\ref{sec:exorem} above). 

The MCMC runs for at most 1,000,000 steps with 20 walkers. It stops whenever the number of steps is smaller than 50$\times$ the largest auto-correlation length of the samples. Table~\ref{tab:MCMC_results} summarizes the results of two different runs, one with all parameters freely varying within uninformed priors (except for $R$). The corner-plot showing the posterior distributions with all parameters freely varying is plotted on Fig.~\ref{fig:MCMC_cornerplot_free}. The synthetic spectrum at the median of these posterior distributions is compared to the observed data on Fig.~\ref{fig:MCMC_model_free}. 
The derived parameters are in good agreement with the initial guess from the grid-search in Section~\ref{sec:gridsearch}, and lead to more reliable confidence regions, with an effective temperature ranging at 1--$\sigma$ within 1555$^{+29}_{-22}$\,K, $\log g$=3.12$^{+0.12}_{-0.09}$, Fe/H=-0.325$^{+0.065}_{-0.045}$\,dex, a super-solar C/O $\sim$0.79$^{+0.01}_{-0.11}$ and a $v\sin i$=31$\pm$5\,km\,s$^{-1}$. Our temperature estimate agrees with the~\citet{Nowak2020} results for the GRAVITY + GPI YJH band data fitted with exo-REM models (1590$\pm$20\,K). Also our derived metallicity and $\log g$ agree with the fit of the GRAVITY-only spectrum. But we found a super-solar C/O where they found it sub-solar. 

The $\log g$$\sim$3.1 found above, as well as by the GRAVITY collaboration~\citep{Nowak2020}, implies a planet mass close to 2\,M$_\text{J}$. It disagrees with the independent dynamical constraints on the mass of the planet $\sim$12\,M$_\text{J}$~\citep{Snellen2018,Lagrange2020,Nowak2020}. We thus try and fix the $\log g$ at a higher value, around 4.18, as suggested by~\citet{Chilcote2017} work, defining a Gaussian prior probability distribution for the $\log g$$\sim$4.18$\pm$0.01. As shown in Table~\ref{tab:MCMC_results}, in this case, the MCMC leads to a larger temperature of 1746$_{-3}^{+4}$\,K, and a smaller C/O of 0.551$\pm$0.002 compatible with a solar value. 

\begin{figure*}\centering
\includegraphics[width=178.6mm]{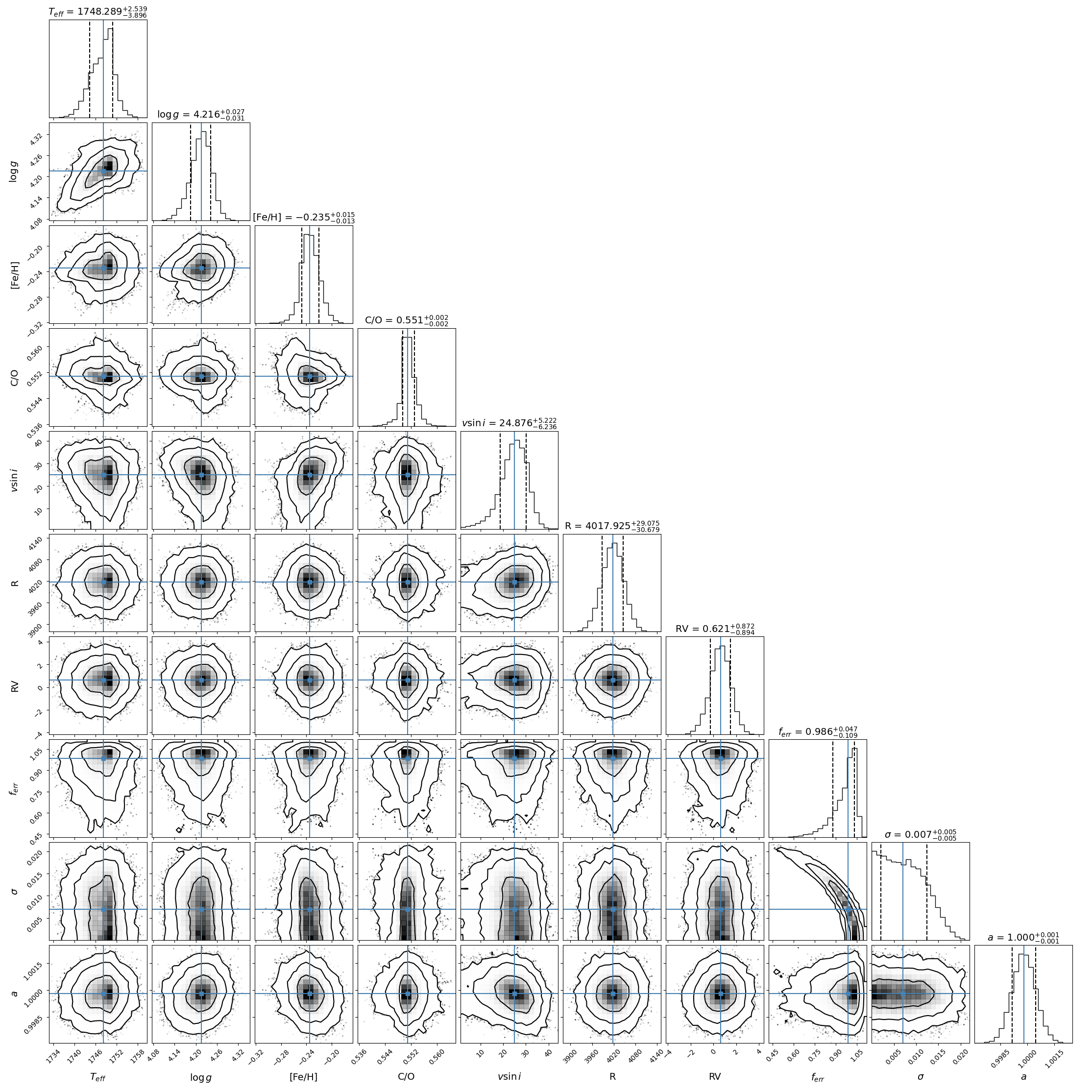}
\caption{\label{fig:MCMC_cornerplot_fixed} Corner plot summarizing the MCMC results for the fit of the $\beta$\,Pic b IR SINFONI spectrum by Exo-REM models with $T_{\rm eff}$ fixed at the~\citet{Chilcote2017}'s value (see text).}
\end{figure*}

\begin{figure*}\centering
\includegraphics[width=178.6mm,clip=true, trim=0 0 0 22]{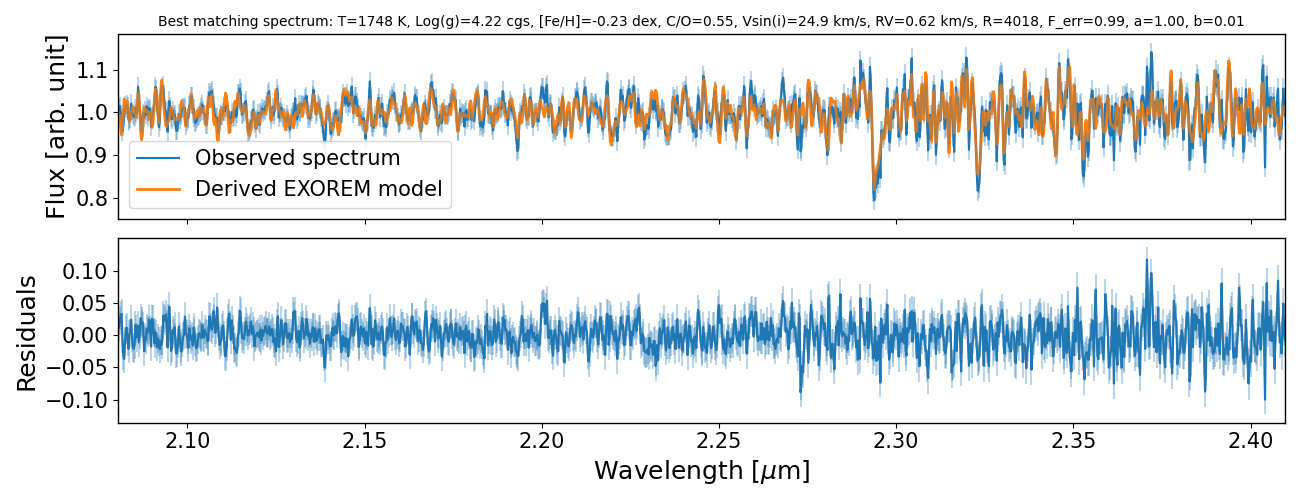}
\caption{\label{fig:MCMC_model_fixed} Plot comparing in the top-panel the $\beta$\,Pic\,b SINFONI spectrum (blue) and the median Exo-REM model (orange) from the MCMC posteriors with $T_{\rm eff}$ fixed at the~\citet{Chilcote2017}'s value (see text). The bottom-panel shows the residuals. The uncertainties assumed for the observed flux are plotted as light-blue vertical lines.}
\end{figure*}

We also tried fixing $T_{\rm eff}$'s prior to~\citet{Chilcote2017}'s value 1724$\pm$15\,K and both $\log g$ and $T_{\rm eff}$. These trials are also shown in Table~\ref{tab:MCMC_results} and are compatible with the fixed-$\log g$ trial. We calculated the Akaike information criterion ${\rm AIC}$=$2k-2\ln{\mathcal L}$ for all models. The maximum likelihood estimator (MLE) model obtained with the $T_{\rm eff}$ fixed minimizes the Akaike information criterion (AIC) and is thus preferred over all other models. The large difference in AIC of $\Delta$AIC$\sim$36 means that the $T_{\rm eff}$--fixed model is 2.5$\times$10$^{-8}$ times more likely than the model with uninformed priors, and 1.4$\times$ more likely than the model with both $\log g$ and $T_{\rm eff}$ fixed. 
This preferred solution has a $T_{\rm eff}$ of 1748$_{-4}^{+3}$\,K, a $\log g$=4.22$\pm$0.03 slightly larger than the Chilcote's value of 4.18, a sub-solar Fe/H=-0.235$^{+0.013}_{-0.015}$\,dex, a solar C/O $\sim$0.551$\pm$0.002, and a $v\sin i$=25$^{+5}_{-6}$\,km\,s$^{-1}$. In all cases, we found a jitter $\sigma$ that is compatible with zero, with a correcting factor of error bars $f_{\rm err}$$\sim$1$^{+0.05}_{-0.11}$ very close to $1$. Both imply that our determination of flux error bars $\sigma_\text{data}$ introduced in~\ref{sec:gridsearch} is reasonable. Fig.~\ref{fig:MCMC_model_fixed} shows our preferred model compared to the observed planet spectrum, and the corner plot of posterior distributions is shown in Fig.~\ref{fig:MCMC_cornerplot_fixed}. 

The radial velocity of the planet is found to be around 0.6$\pm$0.9\,km\,s$^{-1}$ in the Earth reference frame. With a barycentric correction of $8.1$\,km\,s$^{-1}$ this leads to a radial velocity of planet $\beta$\,Pic\,b of 8.7$\pm$0.9\,km\,s$^{-1}$. Compared to $\beta$\,Pic systemic RV of $\sim$20$\pm$0.7\,km\,s$^{-1}$~\citep{Gontcharov2006}, it implies an RV of the planet relative to $\beta$\,Pic's central star of -11.3$\pm$1.1\,km\,s$^{-1}$ at MJD=56910.38. This value agrees at 0.3--$\sigma$ with the predicted RV of the planet at this MJD extrapolating the ephemerides of $\beta$\,Pic\,b's orbital motion from the RV+astro+imaging solution of~\citet{Lacour2021} of -11.6\,km\,s$^{-1}$. It also compares well to the RV measured at high resolution using CRIRES a year earlier, -15.4$\pm$1.7\,km\,s$^{-1}$ at MJD=56,644.5~\citep{Snellen2014}.

\begin{table*}
    \centering
    \caption{\label{tab:MCMC_results} Table summarizing the MCMC results for the run. Unless stated otherwise, all parameters, except $R_{\text{spec}}$, follow uniform priors. We describe the posterior distribution of any parameter by the median and the deviation of the 16th and 84th percentiles to the median. The maximum likelihood estimator (MLE) is given as well, at which point the goodness-of-fit diagnostics are determined.}
    \resizebox{\linewidth}{!}{%
    \begin{tabular}{lc|cc|cc|cc|cc} 
         Parameter & Unit & \multicolumn{2}{|c}{All priors uniform} & \multicolumn{2}{|c}{Prior $\log g$=4.18$\pm$0.01} & \multicolumn{2}{|c}{Prior $T_{\rm eff}$=1724$\pm$15\,K} & \multicolumn{2}{|c}{Prior $\log g$ \& $T_{\rm eff}$} \\
         & & median$\pm$$\sigma_{68.3\%}$ & MLE & median$\pm$$\sigma_{68.3\%}$ & MLE & median$\pm$$\sigma_{68.3\%}$ & MLE & median$\pm$$\sigma_{68.3\%}$ & MLE  \\
         \hline
         $T_\text{eff}$ & K & 1555$_{-29}^{+22}$ & 1525 & 1746$_{-3}^{+4}$ & 1746 & 1748$_{-4}^{+3}$ &  1749 & 1745$_{-2}^{+3}$ & 1746 \\
         $\log g$ & cgs & 3.12$_{-0.09}^{+0.12}$ & 3.24 &  4.185$_{-0.010}^{+0.010}$ & 4.182  & 4.216$_{-0.031}^{+0.027}$ & 4.210 & 4.183$_{-0.010}^{+0.010}$ & 4.190\\
         $[$Fe/H$]$ & dex & -0.325$_{-0.045}^{+0.065}$ & -0.246  &  -0.235$_{-0.011}^{+0.014}$ & -0.226 & -0.235$_{-0.013}^{+0.015}$ & -0.223 & -0.237$_{-0.010}^{+0.013}$ & -0.233 \\
         C/O & --- & 0.79$_{-0.11}^{+0.01}$ & 0.68  &  0.551$_{-0.002}^{+0.002}$ & 0.550  &  0.551$_{-0.002}^{+0.002}$ & 0.550 & 0.551$_{-0.002}^{+0.002}$ & 0.550\\
         $v \sin i$ & km\,s$^{-1}$ & 31.4$_{-5.0}^{+4.6}$ & 33.0 & 27.1$_{-5.4}^{+4.5}$ & 24.9 & 24.9$_{-6.2}^{+5.2}$  & 25.4 & 26.6$_{-5.2}^{+4.4}$  & 28.6 \\
         $R_{\text{spec}}$ & --- & 4018$_{-30}^{+30}$ & 4042 & 4019$_{-30}^{+30}$ & 4024 & 4018$_{-30}^{+30}$ & 4032 &  4019$_{-30}^{+30}$ & 4020 \\
         $v_r$ & km\,s$^{-1}$ &  0.77$_{-0.97}^{+1.05}$  & 1.20 & 0.58$_{-0.89}^{+0.86}$  & 0.30 & 0.62$_{-0.89}^{+0.87}$ & 1.03 & 0.63$_{-0.88}^{+0.84}$ & 0.28 \\
         $f_\text{err}$ & --- & 1.02$_{-0.13}^{+0.05}$ & 1.08 & 0.98$_{-0.12}^{+0.05}$ & 0.98 & 0.99$_{-0.11}^{+0.05}$ & 0.99 & 0.98$_{-0.12}^{+0.05}$ & 1.00\\
         \tablefootmark{$\star$}$\sigma$ & --- & $<0.013$ & 0.002 & $<0.013$ & 0.007 & $<0.012$ & 0.006 & $<0.013$ & 0.005\\
         $a$ & --- & 1.000$_{-0.001}^{+0.001}$ & 1.000 & 1.000$_{-0.001}^{+0.001}$ & 1.000 & 1.000$_{-0.001}^{+0.001}$ & 1.000 & 1.000$_{-0.001}^{+0.001}$ & 1.000\\
         \hline \\
         Goodness-of-fit at MLE \\
         \hline
         $\chi^2$& & & 1345 & & 1375.1 & & 1366.0 & & 1340.5 \\
         $\chi^2_\text{red}$ & & & 1.002 & & 1.029 & & 1.022 & & 1.003 \\
         $\log {\mathcal L}/N_{\text{DoF}}$ & & & 3.274 & & 3.284 & & 3.285 & & 3.281 \\
         AIC & & & -8721.6 & & -8755.6 & & -8758.3 & & -8757.6  \\
         \hline \\
         Diagnostics \\
         \hline
         \tablefootmark{$\dagger$}$N_\text{step}$/max($\tau_\lambda$) & &  \multicolumn{2}{c|}{17} & \multicolumn{2}{c|}{52} & \multicolumn{2}{c|}{50} & \multicolumn{2}{c}{51} \\
         Acceptance rate & & \multicolumn{2}{c|}{0.21} & \multicolumn{2}{c|}{0.31} & \multicolumn{2}{c|}{0.31} & \multicolumn{2}{c}{0.31} \\
         $N_{\text{DoF}}$ & & \multicolumn{2}{c|}{1335} & \multicolumn{2}{c|}{1336} & \multicolumn{2}{c|}{1336} & \multicolumn{2}{c}{1337} \\
         \hline
    \end{tabular}%
    }
    \tablefoot{ \\
    \tablefoottext{$\star$}{For $\sigma$, given the shape of the posterior distribution peaking close to 0 and compatible with 0 at less than 2--$\sigma$, we only give the upper-limit at the 84th percentile.} \\
    \tablefoottext{$\dagger$}{max($\tau_\lambda$) is the maximum auto-correlation time among all varied parameters $\lambda$.}
    }
\end{table*}

\section{Discussion}
\label{sec:discussion}

\subsection{A new estimation of $v\sin i$}
In this work, we found a $v\sin i$ of 25$^{+5}_{-6}$\,km\,s$^{-1}$ in good agreement with the S14's result $v\sin i$=25$\pm$3\,km\,s$^{-1}$ and the most recent~\citet{Landman2023}'s $v\sin i$ estimation of 19.9$\pm$1.1\,km\,s$^{-1}$. The larger confidence region of our measurement is explained by the much lower resolving power of SINFONI ($R$=4030$\pm$30 as determined in Section~\ref{sec:calib}) compared to that of CRIRES ($R$=75,000).  

\begin{figure}
    \centering
    \includegraphics[width=89.3mm]{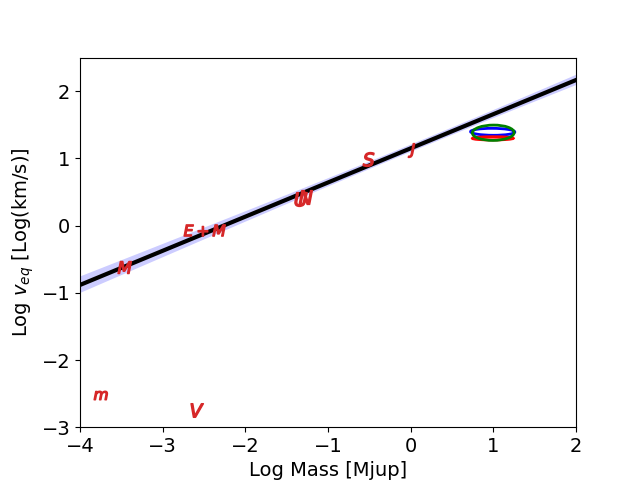}
    \caption{The log-linear relationship between the equatorial velocity and mass of Solar System planets. In blue, possible laws compatible with planets Mars, Jupiter, Saturn, Uranus and Neptune. The ellipses show the $v\sin i$ and mass of $\beta$ Pic b derived in this work (pink) and in S14 (cyan) and~\citet{Landman2023} (red).}
    \label{fig:vsini_mass_ss}
\end{figure}

S14 noticed that given the mass of the planet, the spin velocity of 25\,km\,s$^{-1}$ was too low to comply with the log-linear mass-spin law followed by Solar System objects from Jupiter down to asteroids (e.g. \citealt{Hughes2003}), that would imply a spin velocity of $\sim$50\,km\,s$^{-1}$. Figure~\ref{fig:vsini_mass_ss} shows this relationship for the Solar system planets, with a log-linear law well-fitted to Earth+Moon\footnote{Due to major exchange of momentum between Earth and Moon, the Earth+Moon couple is considered as if the Moon was put back into the Earth and the total momentum conserved~\citep{Hughes2003}.}, Mars, Jupiter, Saturn, Uranus and Neptune mass and equatorial speed, with values taken from~\citet{Hughes2003}. Mercury and Venus are recognized as deviating from this law due to a loss of momentum during their lifetime through tidal interactions  with the Sun~\citep{Fish1967,Burns1975,Hughes2003}. 

\begin{figure}
    \centering
    \includegraphics[width=89.3mm]{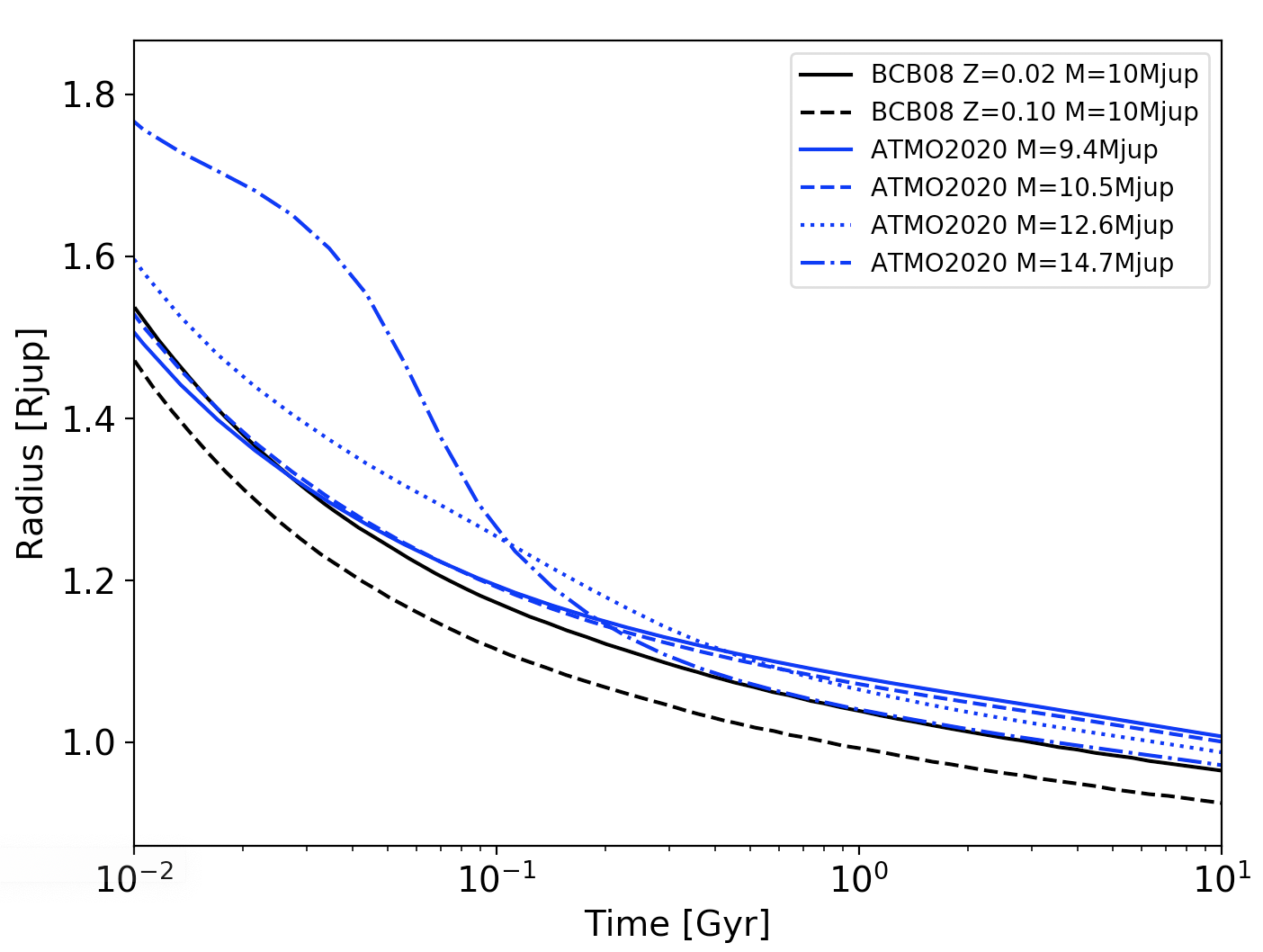}
    \caption{Illustration of evolutionary tracks of equatorial radius through time with ATMO2020 models~\citep{Phillips2020} and Baraffe-Chabrier-Barman~\citep{Baraffe2008} or BCB08 models at different planet mass.}
    \label{fig:evol_radius}
\end{figure}

\begin{figure}
    \centering
    \includegraphics[width=89.3mm]{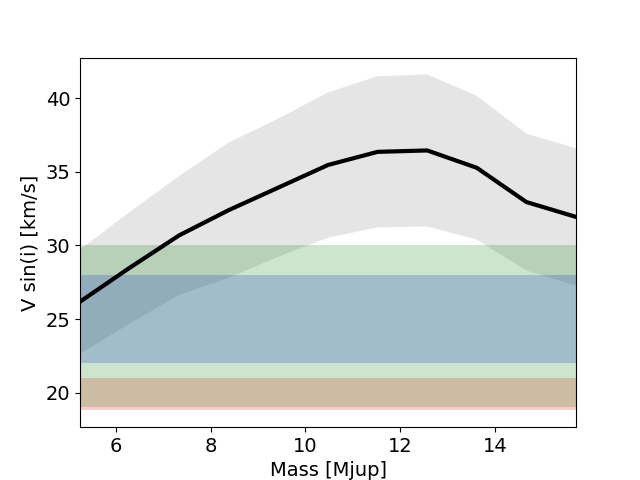}
    \caption{The $v\sin i$ prediction (gray region and black solid line) of the planet for different masses at 23\,Myr, assuming a tilt of $0^\circ$ compared to edge-on. Green and blue area show the $v\sin i$ confidence regions on $\beta$\,Pic\,b based on this work's, S14's (blue) and~\citet{Landman2023}'s (red)} results.
    \label{fig:vsini_mass_evol}
\end{figure}

The difference in $v\sin i$ for $\beta$\,Pic\,b compared to the expected equatorial speed at a mass of $\sim$11\,M$_\text{J}$ is mostly explained by the young age of the planet ($\sim$23 Myr; \citealt{Mamajek2014}). Indeed, the planet is currently contracting from $\sim$1.5\,R$_{\text{J}}$ down to $\sim$1\,R$_{\text{J}}$ (S14, \citealt{Schwarz2016}) and its spin should thus be accelerating. Fig.~\ref{fig:evol_radius} shows the effect of dilatation/contraction through time on the equatorial radius from two different evolution models, ATMO2020~\citep{Phillips2020} and Baraffe-Chabrier-Barman~\citep{Baraffe2008}. According to the conservation of momentum, its spin velocity is expected to increase up to 40$\pm$20\,km\,s$^{-1}$ at 4.5\,Gyr in better agreement with the solar system law. 
Fig.~\ref{fig:vsini_mass_evol} shows the predicted $v\sin i$ at 23\,Myr evolved backward using the ATMO2020 models, down from an equatorial velocity determined by the spin-mass law at 4.5\,Gyr shown in Fig.~\ref{fig:vsini_mass_ss}. 
It can be seen on Fig.~\ref{fig:vsini_mass_evol} that even taking into account contraction there is still a tension between the $v\sin i$ measurements of $\beta$\,Pic\,b and its mass, especially if the mass is contained within 10-14\,M$_\text{J}$. 
The $v\sin i$ and mass overlap only over regions for $v\sin i$$>$22\,km\,s$^{-1}$, with a mass either $<$10\,M$_{\text{J}}$ or $>$14\,M$_{\text{J}}$. These masses are marginally supported by the combined astrometric and RV measurements that rather favour a planet mass within 9-15\,M$_\text{J}$~\citep{Dupuy2019,Nowak2020,Lagrange2020, Vandal2020, Brandt2021, Feng2022}. 

This remaining discrepancy between mass and $v\sin i$ could be explained by the random aspect of moment exchange during planet formation. However, it might also be the hint of a tilt of the planet’s equator compared to its orbital plane. In such case, the projected spin velocity $v\sin i$ is smaller than the true equatorial velocity. A direct compatibility at 1-$\sigma$ of the predicted $v\sin i$ at 23\,Myr for a mass at 11\,M$_\text{J}$ and our measurement 25$\pm$6\,km\,s$^{-1}$ leads to a tilt compared to edge-on $>$15$^\circ$.  

\subsection{A solar C/O}

We derive here a value of the C/O for $\beta$\,Pic\,b that is solar, 0.551$\pm$0.002, while \citet{Nowak2020} and~\citet{Landman2023} found a sub-solar C/O of respectively 0.43$\pm$0.05 and 0.41$\pm$0.04. Forming $\beta$\,Pic\,b in situ along the core accretion scenario~\citep{Pollack1996} would imply a C/O ratio in the atmosphere of the planet largely super-solar $>$0.8, because of the expected abundances of the different gases in the disk from one ice line to the other when assuming a disk with a static composition all through the main phase of planet formation~\citep{Oberg2011}. In this framework, a proposed scenario to reach solar and sub-solar C/O is to consider accretion of icy planetesimals from beyond the ice lines, where most of the H$_2$O and CO$_2$ of the disk is in condensed phase. Alternatively, a solar C/O is more naturally reached by forming either by gravitational instability~\citep{Boss1997} anywhere in the disk, or by core accretion close to the H$_2$O ice line with a moderate planetesimal accretion followed by an outward migration. Given that core accretion is preferred for most compact planetary systems with also terrestrial planets -- as e.g. the Solar system -- and that the $\beta$\,Pic system has at least two planets, with one within 5\,au~\citep{Lagrange2019} plus small km-sized icy bodies~\citep{Ferlet1987,Beust1990,Kiefer2014,Lecavelier2022}, we further consider this scenario as the most likely for forming the $\beta$\,Pic's planets.

In this framework, we can estimate the location of the ice lines compared to $\beta$\,Pic\,$b$ location in the disk. Following~\citet{Oberg2011}, the typical temperature profile in a protoplanetary disk is given by~\citep{Andrews2005,Andrews2007}:

\begin{equation}
    T = T_0 \left(\frac{r}{1\,\text{R}_\star}\right)^{-0.62}
\end{equation}

\begin{figure}
    \centering
    \includegraphics[width=89.3mm]{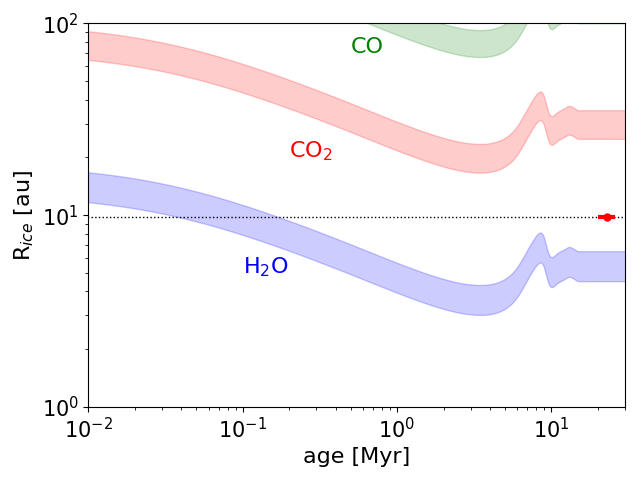}
    \caption{Ice lines locations through time of H$_2$O, CO$_2$ and CO derived for $\beta$\,Pictoris. The location of planet $b$ is marked as a red circle at 23$\pm$3\,Myr and prolonged down to 10 kyr with a black dotted line.}
    \label{fig:icelines}
\end{figure}

Here, $T_0$ is the average temperature in the disk as if it were located at 1\,R$_\star$ from the center of the star. Given the effective temperature of $\beta$\,Pic~\citep{Saffe2021}, we have that $T_0$=8000\,K. Considering the typical evaporation temperatures of H$_2$O, CO$_2$ and CO summarized in Table~\ref{tab:icelines} with $R_\star$=1.7\,R$_\odot$~\citep{Kervella2004}, we derived the radii of the different ice lines around $\beta$\,Pic. They are shown in Table~\ref{tab:icelines} as well. It results that with $a_b$$\sim$9.8\,au~\citep{Lagrange2020}, planet $b$ is located between the H$_2$O (6$\pm$1\,au) and the CO$_2$ (31$\pm$5\,au) ice lines. Fig.~\ref{fig:icelines} shows the variation of the ice lines through time, as derived using Dartmouth12 evolutionary tracks for pre main sequence stars.

\begin{table}
    \centering
    \caption{Ice lines of CO, CO$_2$ and H$_2$O. The values of $T_\text{evap}$ are taken from~\citet{Oberg2011} and the corresponding ice lines radii are calculated as explained in the text. }
    \label{tab:icelines}
    \begin{tabular}{lcc}
      Molecule  & $T_\text{evap}$ (K) & $r_\text{ice}$ (au) \\
      \hline
      CO & 20$\pm$2 & 125$\pm$20\\
      CO$_2$ & 47$\pm$5 & 31$\pm$5 \\ 
      H$_2$O & 135$\pm$15 & 5.7$\pm$1.0 \\
      \hline
    \end{tabular}
\end{table}

Following~\citet{Petrus2021}, our C/O measurement fits well to the~\citet{Nissen2013} planet-C/O to star-Fe/H linear relation. They indeed report a positive correlation between the C/O ratio and the [Fe/H] of stars with an average C/O=0.58$\pm$0.06 at [Fe/H]=0. Elemental abundances and spectral matching agree on a metallicity of $\beta$\,Pic that is solar up to a factor $\sim$2~\citep{Holweger1997}. Our C/O of 0.551$^{+0.003}_{-0.002}$ is in good agreement with the expected solar C/O for a solar metallicity. 
The low metallicity of the planet that we found, -0.235$^{+0.015}_{-0.013}$, is also in agreement with new-generation planetary population synthesis~(NGPPS, \citealt{Schlecker2021}) simulations through core-accretion, on the same range of semi-major axis $\sim$10\,au, especially if the mass of $\beta$\,Pic\,b is larger than 10\,M$_\text{J}$. The correlation between bulk metallicity and planet mass, as obtained through the NGPPS simulations performed around stellar hosts with metallicities ranging from -0.5 to 0.5, can be seen in Fig.~10 of~\citet{Petrus2021}. 

To explain the C/O value found when assuming a static disk composition during planet formation, as in \citet{Oberg2011}, we propose that i) planets b \& c underwent an inner migration during the first Myr of their formation, allowing them to gather large amounts of gas with solar composition; followed by ii) an outward migration, with planet b \& c in 7:1 mean motion resonance leading them to reach their actual location (Beust et al., priv. comm.). This scenario would allow avoiding a fine-tuned icy planetesimal accretion within the planet to reach an almost perfect solar C/O. 

Alternatively, there exists a scenario that do not require invoking outward migration of $\beta$\,Pic\,b. Considering non-static disk composition, \citet{Molliere2022} found that pebble evaporation and the dilution of water and CO in-between the H$_2$O and CO iceline (Fig.~6 in \citealt{Molliere2022}) can lead to a nearly stellar C/O ratio in the circumstellar gas in about 1 Myr. That could enable in-situ formation of $\beta$\,Pic\,b provided most of its gas was not accreted before 1\,Myr. 

To conclude, a solar C/O for $\beta$\,Pic\,b is challenging for planet formation models. Interestingly, it fits in the mass-C/O relation obtained by~\citet{Hoch2022}. They found a clear threshold at 4\,M$_\text{J}$ beyond which imaged planet mostly have C/O consistent with solar $\sim$0.55, while below 4\,M$_\text{J}$, transiting planet have C/O with any values from 0.2 to 2.0. This threshold has been interpreted as distinguishing two main formation pathways for planets either less massive than 4\,M$_{J}$ and or either more massive than 4\,M$_{J}$. This 4-M$_\text{J}$ threshold was already discovered in the distribution of stellar metallicities among massive Giant exoplanets with a similar interpretation~\citep{Santos2017}. Therefore, the solar C/O of $\beta$\,Pic\,b might be the sign of a formation scenario that differ from the usual core accretion scenario invoked for close-in less massive planets, either still by core accretion but considering perhaps a distinct pathway for planets formed at large separation that did not undergo inward migration, or either more simply by gravitational instability.

\section{Conclusion}
\label{sec:conclusion}

In this study we have derived a new infrared spectrum of the young giant planet $\beta$\,Pic\,b observed with SINFONI. Doing so, we have shown that the actual spectral resolving power of SINFONI in the K-band at a spaxel-resolution of 12.5$\times$25\,mas$^2$ is $\sim$4000. Then, using a novel method of stellar halo removal, we have been able to directly extract the spectrum and the molecular lines of the planet without the need of using molecular mapping techniques. We have fitted the spectrum to models from the forward--modeling Exo-REM library. This led to different results, depending on assumptions on the planet mass and radius: 
\begin{itemize}
    \item without any prior constraints, we have obtained $T_\text{eff}$=1555$^{+22}_{-29}$\,K at a $\log g$=3.12$_{-0.09}^{+0.12}$, with sub-solar metallicity of -0.325$^{+0.065}_{-0.045}$\,dex and a super-solar C/O=0.79$^{+0.01}_{-0.11}$;
    \item assuming a  prior on the $T_{\rm eff}$=1724$\pm$15\,K based on~\citet{Chilcote2017} independent photometric characterisation, we have rather found higher $T_\text{eff}$=1748$^{+3}_{-4}$\,K, with again sub-solar metallicity of -0.235$^{+0.015}_{-0.013}$\,dex and a now solar C/O=0.551$\pm$0.002.
\end{itemize}

Our preferred parameters are those derived imposing $T_{\rm eff}$=1724$\pm$15\,K, as it better reflects the gravitational mass and the geometric radius derived independently using photometry. 
We find a projected rotation speed of $\beta$\,Pic b's equator of 25$^{+5}_{-6}$\,km\,s$^{-1}$ agreeing with the 25$\pm$3\,km\,s$^{-1}$ found by~\citet{Snellen2014} at high spectral resolution with CRIRES and the most recent CRIRES+~\citet{Landman2023}'s $v\sin i$=19.9$\pm$1.1\,km\,s$^{-1}$. However, our measurement of a solar C/O is in stark contrast with the sub-solar C/O$\sim$0.45 obtained from GRAVITY spectra~\citep{Nowak2020} and by atmospheric retrieval of CRIRES spectra in~\citet{Landman2023}.

Conversely to the conclusions of~\citet{Nowak2020},  with a C/O for the star also close to 0.55, the stellar C/O disfavours the scenario of icy-planetesimals injection. It would indeed imply fine-tuning of the planetesimals accretion flux and of the duration of the phenomenon so that it reaches a value close to the stellar C/O. Such value of $\beta$\,Pic\,b's C/O is on the other hand in better agreement with in situ gravitational instability scenario, or formation next to the H$_2$O ice line followed by migration. This latter scenario could be supported by the proximity of the planets b \& c orbital periods to a 7:1 resonance, whose orbital parameters may have significantly evolved during the first Myrs of the system.

Finally, we measure a radial velocity of $\beta$\,Pic\,b relative to the central star of -11.3$\pm$1.1\,km\,s$^{-1}$ at MJD=56,910.38. It agrees at 0.3--$\sigma$ with the ephemerides for the orbit of this planet based on the current knowledge of the system from~\citep{Lacour2021}. This tends to confirm the current estimation of the orbits and mass of the $\beta$\,Pic's planets, and add a new RV point for future dynamical characterisation of the system. 

The presented work shows that ground-based infrared medium-resolution spectroscopy with even a modest resolving power of 4000 -- close to that of the JWST/MIRI-MRS -- without the need of doing molecular mapping and with a careful treatment of wavelength calibration, as well as star halo and telluric lines subtraction at any wavelength, can allow deriving key properties of imaged exoplanets, including equatorial rotation velocity. Proper determination of the atmospheric parameters though still requires independent priors, such as on $T_{\rm eff}$ or $\log g$, due to fit degeneracy of the unresolved spectral lines. 

\begin{acknowledgements}
We are thankful to Alain Smette for his careful and constructive reading of the manuscript that helped improving the analysis significantly. This work was based on observations collected at the European Southern Observatory under ESO programme 093.C-0626(A). It was granted access to the HPC resources of MesoPSL financed by the Region Ile de France and the project Equip\@Meso (reference ANR-10-EQPX-29-01) of the programme Investissements d’Avenir supervised by the Agence Nationale pour la Recherche. This project has received funding from the European Research Council (ERC) under the European Union's Horizon 2020 research and innovation programme (COBREX; grant agreement n° 885593). This work was also funded by the initiative de recherches interdisciplinaires et stratégiques (IRIS) of Université PSL "Origines et Conditions d'Apparition de la Vie (OCAV)". FK also acknowledge funding from the Action Incitative Exoplanètes de l'Observatoire de Paris. 
\end{acknowledgements}

\bibliographystyle{aa}
\bibliography{main_biblio}

\begin{appendix}
\onecolumn
\section{MCMC run with all free parameters}
\begin{figure}[hbt]\centering
\includegraphics[width=178.6mm]{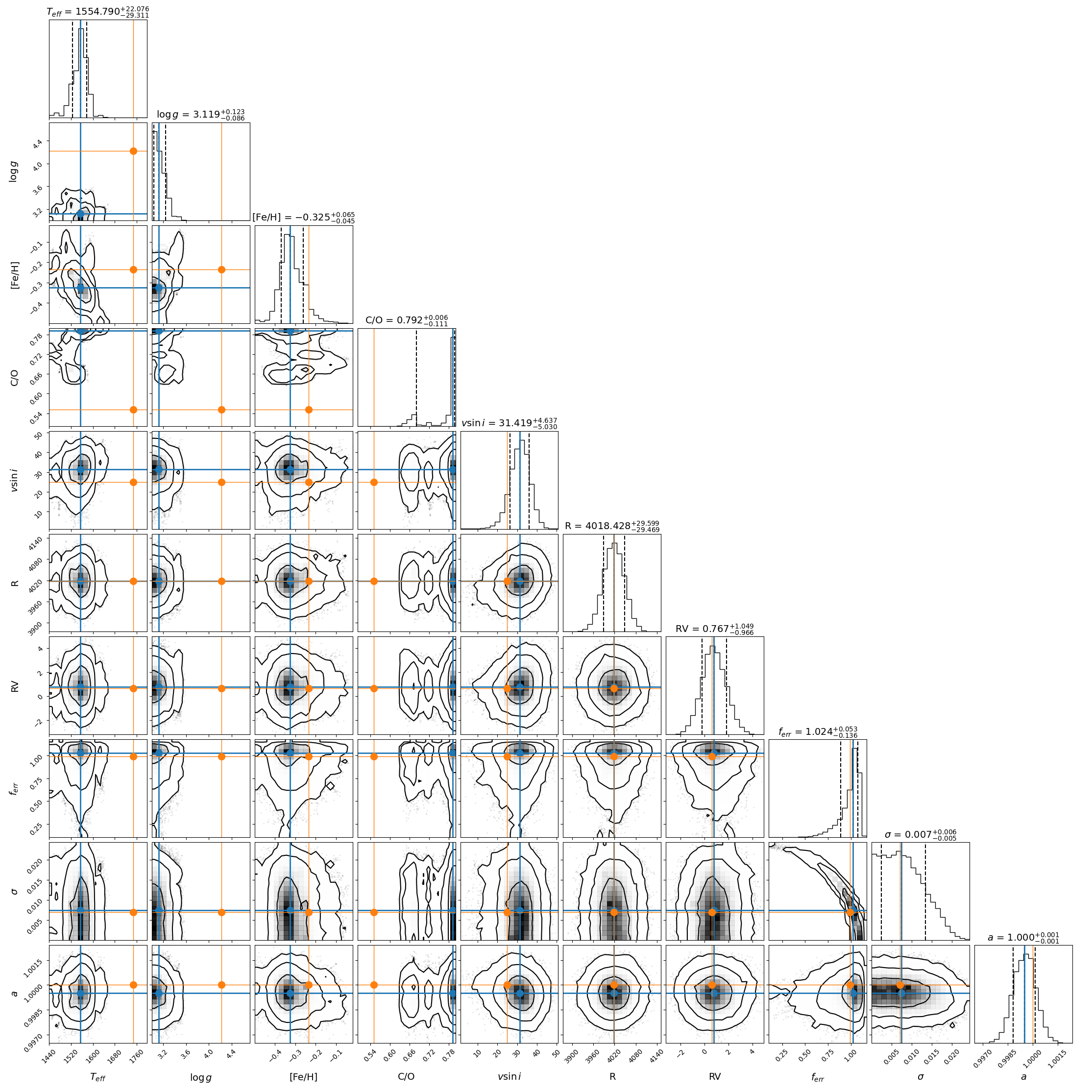}
\caption{\label{fig:MCMC_cornerplot_free} Corner plot summarizing the MCMC results for the fit of the $\beta$\,Pic b IR SINFONI spectrum by Exo-REM models with all free parameters. The blue lines and dot show the median parameters ; the orange lines and dot are those found for the preferred solution (see Section~\ref{sec:mcmc}).}
\end{figure}

\begin{figure}[hbt]\centering
\includegraphics[width=178.6mm,clip=true,trim=0 0 0 22]{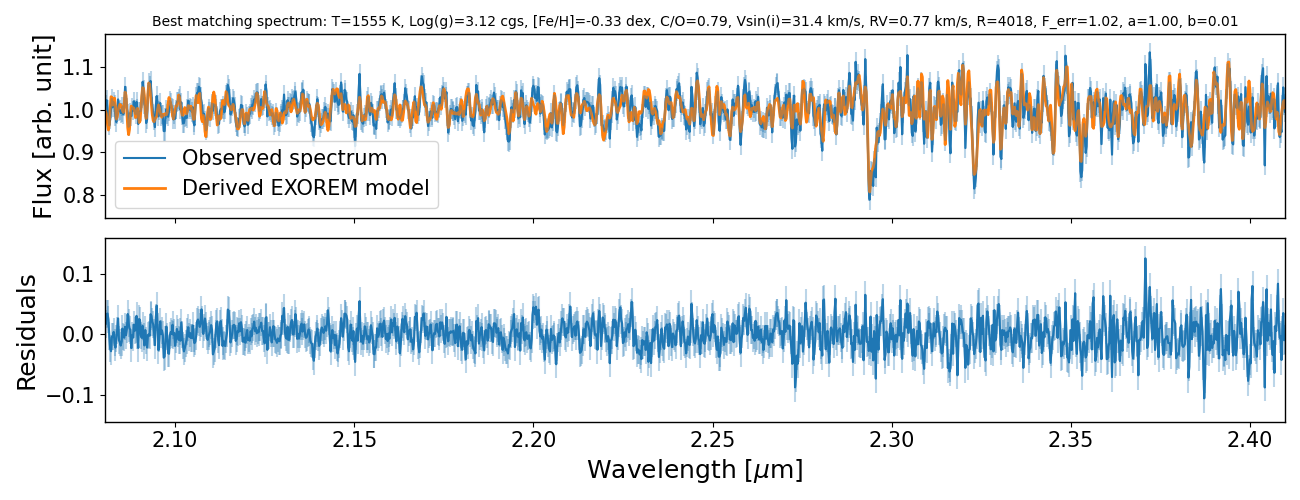}
\caption{\label{fig:MCMC_model_free} Plot comparing the $\beta$\,Pic\,b SINFONI spectrum (blue) and the median Exo-REM model (orange) from the MCMC posteriors (Fig.~\ref{fig:MCMC_cornerplot_free}). The lower-panel shows the residuals. The uncertainties assumed for the observed flux are plotted as light-blue vertical lines.}
\end{figure}
\end{appendix}

\end{document}